\documentclass[12pt]{article}
\pdfoutput=1
\usepackage[square,comma,numbers,sort&compress]{natbib}
\usepackage{graphicx,epstopdf,amssymb,amsfonts,amsmath,amsthm,array,
mathrsfs,amscd}
\usepackage{hyperref}
\usepackage[vcentermath]{youngtab}
\newcommand{\pbs}[1]{\let\temp=\\#1\let\\=\temp}
\numberwithin{equation}{section}
\def\be{\begin{equation}}\def\ee{\end{equation}}
\def\cvp{\raise 2pt\hbox{,}} 
 \def\tr{\mathop{\rm tr}\nolimits}

 \def\d{{\rm d}}\def\nn{{\cal
N}} 
 \def\uN{\text{U}(N)}\def\uK{\text{U}(K)}

\def\ls{\ell_{\text s}}
\def\la{\lambda}\def\La{\Lambda}
\def\A{\mathscr A}
\def\plb#1#2#3{{\it Phys.\ Lett.\ }{\bf B #1} (#2) #3}
\def\npb#1#2#3{{\it Nucl.\ Phys.\ }{\bf B #1} (#2) #3}
\def\npps#1#2#3{{\it Nucl.\ Phys.\ Proc.\ Suppl.\ }{\bf #1} (#2) #3}
\def\prl#1#2#3{{\it Phys.\ Rev.\ Lett.\ }{\bf #1} (#2) #3}
\def\jhep#1#2#3{{\it J. High Energy Phys.\ }{\bf #1} (#2) #3}
\def\prd#1#2#3{{\it Phys.\ Rev.\ }{\bf D #1} (#2) #3}

\def\atmp#1#2#3{{\it Adv.\ Theor.\ Math.\ Phys.\ }{\bf #1} (#2) #3}
\def\cmp#1#2#3{{\it Comm.\ Math.\ Phys.\ }{\bf #1} (#2) #3}
\def\pr#1#2#3{{\it Phys.\ Rep.\ }{\bf #1} (#2) #3}

\def\ijmpa#1#2#3{{\it Int.\ J.\ Mod.\ Phys.\ }{\bf A #1} (#2) #3}
\def\mpla#1#2#3{{\it Mod.\ Phys.\ Lett.\ }{\bf A #1} (#2) #3}

\def\imath#1#2#3{{\it Invent math }{\bf #1} (#2) #3}

\begin{document}
%
%
{\pagestyle{empty}
\parskip 0in
\

\vfill
\begin{center}
{\LARGE Gauge Theories, D-Branes and Holography}



\vspace{0.4in}

Frank F{\scshape errari}
\\
\medskip
{\it Service de Physique Th\'eorique et Math\'ematique\\
Universit\'e Libre de Bruxelles and International Solvay Institutes\\
Campus de la Plaine, CP 231, B-1050 Bruxelles, Belgique}
\smallskip
{\tt frank.ferrari@ulb.ac.be}
\end{center}
\vfill\noindent

Based on a generalization of the string theoretic concept of D-brane probe, we propose a new approach to large $N$ gauge theories which makes the holographic properties manifest. For any gauge theory, we define from first principles an effective action for a fixed number of ``probe'' D-branes in the presence of $N$ ``background'' D-branes on which the gauge theory lives. This effective action is shown to encode all the information about the large $N$ gauge theory. The analysis of the planar diagram expansion which computes the effective action yields a simple and generic mechanism explaining the emergence of
holographic space dimensions: the probe D-branes move in a higher dimensional dual holographic space-time.
The construction yields a new perspective on the notion of bulk space-time locality and draws unexpected links with some aspects of the 't~Hooft Abelian projection ideas. It also provides a new non-perturbative approximation scheme, able to capture both the weak and strong coupling regimes. We sketchily illustrate the basic ideas on a few examples, including the pure four dimensional Yang-Mills theory.

\vfill

\medskip
%
\begin{flushleft}
October 25th, 2013
\end{flushleft}
\newpage\pagestyle{plain}
\baselineskip 16pt
\setcounter{footnote}{0}

}

\section{\label{s1Sec} General presentation}
\subsection{Introduction\label{IntroSec}}

One of the great lesson of the gauge/string correspondence, and in particular of its explicit realization in the AdS/CFT context \cite{AdSCFT}, is that string theory with all its startling features must actually be contained in ordinary flat-space quantum gauge theories. From the point of view of the gauge theory, extra dimensions, quantum gravity, strings, D-branes, etc., should somehow emerge from strong quantum mechanical effects, yielding a dual bulk holographic description which can be extremely useful to study the strong coupling gauge theory physics. This picture is believed to be valid for essentially any matrix gauge field theory, including non-supersymmetric pure Yang-Mills or QCD \cite{polgen}. Unfortunately, convincing explicit evidence exists only for highly supersymmetric models, like the four-dimensional $\nn=4$ gauge theory, for which direct constructions in the framework of standard superstring theory are available and under control. In more generic situations, one is desperately lacking a simple picture which would allow to understand the emergence of the holographic bulk geometry. 

Developing such a picture may seem very challenging, but the potential benefits could be great. On the one hand, the most difficult aspects of the strongly coupled physics of gauge fields can often be elegantly and straightforwardly understood in the holographic framework. For example, it is reasonable to think that a proof of the permanent confinement of quarks by the strong nuclear force would follow in a natural way from a holographic picture of the pure Yang-Mills theory in four dimensions. On the other hand, the gauge theoretic description of the holographic space-time allows, in principle, to address hard questions in quantum gravity in a perfectly well-defined framework, from first principles. For example, one should be able to work out a manifestly unitary picture of the process of black hole formation and evaporation purely in gauge theoretic terms. In short, a better understanding of the origin of holography may bring crucial insights into some of the deepest open problems in both gauge theory and quantum gravity.

Of course, the conceptual difficulties hindering a direct understanding of the emergence of a dual holographic bulk space-time from a pre-geometric, gauge theoretic model are manifold. Some of the issues that the author of the present paper have found particularly difficult or even puzzling, and on which we shall be able to shed some light in this paper, are the following. 

\medskip

\noindent\emph{The bulk geometry must fluctuate}

Let $N$ be the number of colors in the gauge theory, i.e.\ the size of the matrix fields. The emerging space must be classical in the large $N$ limit, but it must fluctuate when $1/N$ corrections are taken into account. It is hard to find candidates that satisfy both requirements. For example, the renormalization group (RG) scale is a natural candidate for one of the coordinates. At infinite $N$ and in conformal field theories, it is well-known that the RG scale can indeed be unambiguously identified with the radial coordinate in AdS (see e.g.\ \cite{verlinde} and references therein). However, there is no obvious definition of a notion of ``fluctuating RG scale.'' At a fundamental level, this identification is thus probably a red herring, since it does not make sense beyond the leading large $N$ approximation. 

\smallskip

\noindent\emph{The emerging holographic bulk is not always a moduli space}

In all the known explicit examples of holographic duals, the emerging space is related to the moduli space of the underlying supersymmetric model. This is true both for the Matrix Theory \`a la BFSS \cite{BFSS} and for AdS/CFT. For example, in the $\nn=4$ theory, the six dimensional moduli space of vacua is naturally associated with the six emerging dimensions of the $\text{AdS}_{5}\times\text{S}^{5}$ geometry. This relationship is actually instrumental in the way the duality is traditionally introduced, when one starts with a stack of D3-branes that can freely move in ten dimensions and then take the near horizon limit. However, this correspondence between moduli spaces and emergent holographic dimensions should not be fundamental. 
This is particularly clear if one believes that the closed string holographic description of gauge fields applies to models with no supersymmetry, no scalar field and thus no moduli. Of course, with no moduli space or near horizon limit to start with, it is a priori unclear how holographic dimensions could emerge.

\smallskip

\noindent\emph{Emerging dimensions are typically not large}

Unlike in highly supersymmetric models with a moduli space, it is not expected that the emerging holographic dimensions will be ``large'' for more   generic or realistic gauge theories. In general, one has to deal with a ``stringy'' holographic geometry. For example, in the pure Yang-Mills theory, there is only one available scale $\Lambda$, which is dynamically generated, and all scales in the problem are naturally of the order of $\Lambda$. In particular, the curvature in the emerging dimensions and the string length, which governs the masses of excitations, are likely to be of the same order of magnitude. A classical description is of course still possible, but it has to involve a non-local action, including all the $\alpha'$ corrections, or even include an infinite number of degrees of freedom associated with the excited string states. This seems to imply that the holographic description cannot be simple or even tractable. Only in some special circumstances, like at large 't~Hooft's coupling in the $\nn=4$ theory, does a local derivative expansion of the action, and a standard Einstein's description of gravity, can give an accurate description. One may hope that such a simplification also occurs, to some level of accuracy, in the infrared region of the emerging space for the pure Yang-Mills theory, which is associated with the strongly coupled dynamics. 

\smallskip

\noindent\emph{Local bulk physics is not observable and thus not gauge invariant}

This last point is probably the deepest and most puzzling. The difficulty is that the holographic bulk coordinates, or actually any other local quantity in the bulk, are not observable. This must be a fundamental feature of the holographic  space-time, tied to the fact that there is no local observable in quantum gravity. It seems to the present author that, from the gauge theory point of view, the only consistent interpretation of this basic property of quantum gravity is that \emph{the construction of a local holographic bulk description cannot be a gauge-invariant procedure.} This fact, which we believe is central, seems to have been largely overlooked in the literature. For example, previous interesting attempts to reconstruct the bulk space-time from the gauge theory that we know of barely mention the problem of gauge-fixing in a footnote \cite{reconstruct1} or ignore it completely \cite{reconstruct2}. The one exception is \cite{insightbrane}, where some insightful comments are made. One of the striking consequence of the gauge dependence is that the emergent bulk coordinates cannot be gauge invariant operators (either local or non-local) in the gauge theory. Having to deal with non gauge invariant objects is of course perfectly consistent and not, by itself, a flaw. The gauge potential is a startling example of an object which is not gauge invariant but does play a fundamental r\^ole in the understanding of weakly coupled gauge theories. The non gauge invariant local description of the holographic bulk space-time is potentially as useful as the gauge potential, but most likely in the opposite, strongly coupled, regime. This being said, it is a priori quite non-trivial to guess which kind of non gauge invariant approach could be associated with the holographic description of the gauge theory. 

\medskip

The aim of the present paper is to propose a route from which a holographic description of gauge theories naturally emerges and which provides insights into all the above issues, particularly on the nature of bulk locality. The idea is that a powerful way to understand the emergence of the holographic dimensions is to study how the fundamental string theory concepts of D-branes and non-abelian D-brane effective actions are built in field theory.
Our construction does not rely at all on supersymmetry or even on the existence of an underlying fundamental string theory set-up. Of course, when this is the case, for example if the gauge theory under study is the $\nn=4$ super Yang-Mills theory, then a direct link between our considerations and the known holographic description can be made. The present work can actually be viewed as a synthesis of the results obtained during the last year by the author and its collaborators by studying some specific examples in string theory \cite{fer1, fer2, fer3, fer4, fer5}. However, possibly the most important contribution of the research presented here is to break up with familiar supersymmetric set-ups in standard string theory and develop tools to understand holography in contexts much wider than those that have been explored before. This is certainly a long and dangerous leap forward, but the benefits are worth the risk. Moreover, the big picture we get seems to be perfectly consistent and draw unexpected links with seemingly unrelated approaches to gauge theory. 

Let us now briefly introduce some of our basic results.

\subsection{Overview of the results \label{genDpSec}}

The basic tool we shall develop is a generalization of the string theoretic notion of D-brane, particularly of D-brane probe, to any gauge theory, including the pure Yang-Mills model, in any number $p+1$ of space-time dimensions. One may even start from a full-fledged open string field theory, from which a standard gauge field theory is obtained in the low energy limit. However, for concreteness, we shall mainly deal with ordinary gauge theories with a unitary gauge group $\uN$.

The key object of our construction is an action $\A_{K,N}$ which is interpreted as representing the effective action for $K$ D$p$-branes in the presence of $N$ D$p$-branes on which the original gauge theory lives. The action $\A_{K,N}$ will be defined in purely gauge-theoretic terms and from first principles, with no reference to string theory. A particularly interesting special case is to take $N\rightarrow\infty$ at fixed $K$. The action $\A_{K,\infty}$ then describes the dynamics of $K$ ``probe'' branes in the presence of a very large number of ``background'' branes (this terminology will be explained shortly).

Unless explicitly stated otherwise, the D-branes we refer to will always be D$p$-branes, all ``parallel'' and of the same types. Systems made of branes of different types and dimensionality can also be introduced, as we shall briefly explain, but they won't be our main concern in this work.

The action $\A_{K,N}$ has many very interesting properties.

\noindent (i) It has a $\uK$ gauge symmetry.

\noindent (ii) Its large $N$, fixed $K$, expansion is of the form
\be\label{AKNexp} \A_{K,N} = \sum_{h\geq 1,\, g\geq 0} N^{2-h-2g}A_{K}^{(h,g)}\, \ee
where $A_{K}^{(h,g)}$ gets contributions from Feynman diagrams with $h$ loops of $\uK$ indices which can be mapped to worldsheets of genus $g$ having $h$ boundaries on the $K$ ``probe'' D-branes. In particular, using the simplified notation 
\be\label{AKdef} A_{K}=A_{K}^{(1,0)}\, ,\ee
we get
\be\label{leadinglargeN} \A_{K,N}\simeq N A_{K}\ee
to leading order. The $\uK$-invariant action $A_{K}$ is independent of $N$ and gets contributions from planar diagrams with only one loop of $\uK$ indices, corresponding to a unique worldsheet boundary on the probe branes. It has a single-trace structure.

\noindent (iii) Let $Z_{N}$ be the partition function of the original gauge theory with gauge group $\uN$. As is well-known, the large $N$ expansion of its logarithm has the form
\be\label{ZNexp} \ln Z_{N} = -\sum_{g\geq 0}N^{2-2g} F^{(g)}\, ,\ee
where $F^{(g)}$, the genus $g$ free energy, is given by a sum over genus $g$ Feynman diagrams. There is a fundamental relation between the planar free energy $F^{(0)}$ and the probe brane action,
\be\label{fundrel} F^{(0)} = \frac{1}{2K} A_{K}^{*}\, .\ee
The star-spangled notation $A_{K}^{*}$ means that the probe brane action is evaluated on-shell. In other words, $A_{K}^{*}$ is the value of the action $A_{K}$ on the solution of its classical equations of motion. Since the partition function $Z_{N}$ may be defined by adding sources for any gauge-invariant operator in the theory, the equation \eqref{fundrel} shows that \emph{the probe brane action $A_{K}$ contains all the information about the planar limit of the original gauge theory.} It is actually not useful, from this point of view, to deal with a non-abelian action. Indeed, the left-hand side of \eqref{fundrel} obviously does not depend on $K$ and thus, on-shell, all the D-brane actions are trivially related. Defining
\be\label{Adef} A = A_{1}\, ,\ee
we must have
\be\label{AKAonshell} A_{K}^{*} = K A^{*}\ee
and thus
\be\label{fundrel2} \boxed{F^{(0)} = \frac{1}{2}A^{*}\, .}\ee
We shall write down more general relations between the partition function $Z_{N}$ and the actions $\A_{K,N}$, valid to all orders in the $1/N$ expansion or even at finite $N$.

\noindent (iv) In the cases where the gauge theory under study has a known holographic supergravity dual, and when $N\gg K$, then the action $\A_{K,N}$ matches the non-abelian D-brane action for $K$ D-branes moving in the dual supergravity background. This is why we have used the terminology of ``probe'' branes above, even though the action $\A_{K,N}$ is defined purely in terms of the gauge theory, with \emph{no} a priori reference to a holographic dual geometry that could be probed. 
Note that the full supergravity solution can be read-off from $A_{K}$, by comparing various terms in this action with the known form of the non-abelian D-brane action in general supergravity background \cite{Myers}. This idea was used successfully in \cite{fer1,fer2,fer3,fer4,fer5}. In this sense, $A_{K}$ contains all the information about the background. 
Note also that the relation \eqref{fundrel2} provides a new holographic prescription to compute the partition function and thus the correlation functions of gauge-invariant operators. This new prescription is based on evaluating the on-shell D-brane probe action, whereas the standard recipe is based on evaluating the on-shell supergravity action. The equality \eqref{fundrel2} implies that the on-shell supergravity action should always be equal to half the on-shell D-brane action for a single D-brane probe.

\noindent (v) One of our main result will be to show that the sum over an infinite number of planar loop diagrams computing the probe brane dynamics  \emph{naturally produces an action depending on a certain number of scalar fields in the adjoint representation of $\uK$.} This is true even for gauge theories, like the pure Yang-Mills, which do not contain any elementary scalar fields in their Lagrangian. These scalar fields in the D-brane action are interpreted, as usual, as corresponding to the coordinates of a target space in which the D-branes move. In other words, the action $A_{K}$ always describes the classical motion of probe branes in a higher dimensional space. This space is the emergent holographic geometry dual to the original gauge theory. Even more generally, we shall show that the full description of the probe brane dynamics involves in principle an infinite number of independent field theoretic degrees of freedom. This is consistent with the existence of an infinite tower of open string states living on the probe brane worldvolume. Note that this is true even though the original D-brane theory is a standard gauge theory with a finite number of fields. Eventually, we get a full open string field theory description of the probe branes moving in the emergent holographic geometry sourced by the background branes.

\noindent (vi) The action $\A_{K,N}$ and thus, in particular, the leading large $N$ action $A_{K}$, is gauge-dependent; it depends on a particular partial gauge-fixing procedure which is required to define it. \emph{This gauge-fixing procedure is non-standard and conceptually different from a background field gauge approach.} It was recently worked out in \cite{ferequiv} (see also \cite{Schaden,Shamir}) and relies on an equivariant BRST formalism implying the presence of quartic ghost couplings in the tree-level Lagrangian. This crucial gauge-dependence of the construction is directly related to, and provides a new enlightening point of view on, the problem of bulk locality. It yields a mathematically precise notion of fuzziness which is totally unrelated to the usual quantum gravity fluctuations at the Planck scale. It might even provide a precise formulation of the notion of black hole complementarity \cite{complementarity}. Probe brane actions defined using distinct partial equivariant gauge-fixing terms can thus yield very different local holographic space-time descriptions of the same physics.
Of course, the physical observables themselves must be gauge-independent. This is encoded in the fundamental relation \eqref{fundrel}, which shows that the on-shell action $A_{K}^{*}$ is always gauge-independent, even though $A_{K}$ itself is not. This is a rather non-trivial mathematical property, that follows from the results in \cite{ferequiv} and which relies on the presence of the quartic ghost terms. Another very interesting aspect of the partial gauge-fixing procedure is to make an unexpected link with ideas akin to those developed in the context of the Abelian projection scenario invented by 't~Hooft \cite{tHooftAP} and which have been thoroughly studied over the years, in particular on the lattice (see e.g.\ \cite{APreview} for a review). For example, the emerging dimensions turn out to be quite analogous to condensates that have been studied on the lattice to explain Abelian dominance in QCD.

\noindent (vii) The planar Feynman diagrams computing the probe action $A_{K}$ have a mixed vector/matrix model structure. This suggests a new non-perturbative approximation scheme in gauge theory, which we call the ``bare bubble approximation.'' In this approximation, one resums exactly, already at leading order, an infinite set of planar loop diagrams, corresponding to worldsheets of all relevant topologies. A startling property is that this partial resummation of diagrams captures the emergence of the  holographic dimensions. One may thus hope that it is able to capture as well some important features of the non-perturbative, strongly coupled physics, including in non-supersymmetric contexts. The existence of this new approximation is one of the most exciting aspect of the whole story. It is one-loop exact and thus always reliable at weak coupling, and at the same time provides a new analytical window on strongly coupled gauge theory physics. When applied to the simplest, yet quite non-trivial, matrix gauge theory, the one matrix model in zero dimension, it yields the free energy with an unreasonable accuracy (better that $3\%$!) for \emph{all} values of the 't~Hooft coupling $\la$ \cite{onemmbranes}. In particular, it reproduces the main qualitative features of the strong coupling expansion, yielding the correct asymptotic form at $\la\rightarrow\infty$ and a large $\la$ expansion in powers of $1/\sqrt{\la}$.

\subsection{Plan of the paper}

In Section 2, we introduce the notion of D-brane probe actions in general gauge theories, describe their expected large $N$ properties and provide a few simple explicit examples. In Section 3, we work out the main features of the Feynman diagram expansion computing the D-brane actions and explain how this inevitably leads to a classical description of the probe D-branes moving in an emergent holographic background. In the case of the pure Yang-Mills theory in four dimensions, this picture yields one emerging dimension and thus a five dimensional holographic bulk geometry. In Section 4, we add a few important ingredients to the discussion: we explain that the construction really yields a mapping between states and geometries; we show that the D-brane action encodes the full large $N$ physics, i.e.\ the generating functional of planar gauge-invariant correlation functions can be obtained from it; we also explain that the probe D-brane theory is, strictly speaking, an open string field theory, with an infinite number of independent fields living on it; and finally we discuss consequences of this fact for the concept of emerging dimensions. In Section 5, 
we introduce and discuss the bare bubble approximation. In Section 6, we complete the definition of the probe brane actions by discussing the problem of gauge-fixing. This is a crucial and subtle part of the construction, based on a technology developed recently in \cite{ferequiv}. In Section 7, we discuss the physical consequences of our work for the problem of bulk space-time locality and also reveal unexpected links between D-brane physics, holography and the Abelian projection ideas. Finally, in Section 8 we summarize our main results and briefly discuss some of the many possible future applications. 

We shall illustrate our ideas on several new examples below, including the pure Yang-Mills theory in four dimensions. However, our emphasis is on the general framework more than on the study of specific examples, which is only sketched. Much more details on particular models will be given in forthcoming publications, starting in \cite{onemmbranes}.

\section{D-branes in gauge theories}\label{Dbranesec}

In this section, we introduce the notion of D-branes and most particularly of D-brane probes in any gauge theory. We take a simplified point of view, discarding, for the moment, the problem of gauge-fixing. This is useful because several important aspects of the construction, discussed in Sections \ref{diagsec}--\ref{bareapsec}, are not modified in any essential way by the gauge-fixing procedure. The gauge-fixing itself, which is actually essential for the consistency of the whole picture, is discussed in Section \ref{gfsec}.

\subsection{D-branes}

We consider a gauge theory of gauge group $\uN$ with only adjoint fields. The space-time manifold is $\mathbb R^{p+1}$. The fields are thus all $N\times N$ Hermitian matrices. For example, we could study the pure Yang-Mills gauge theory, or any supersymmetric generalization with adjoint fields. Note that working with other gauge groups and/or considering fields in other representations, like fundamental quarks, or putting the model on a more general space-time manifold, is not expected to change the discussion in any essential way.

We denote the matrix fields of the model by the generic notation $M^{a}_{\ b}$, where the indices $a$ and $b$ run from $1$ to $N$. In general, $M$ represents a collection of fields, which typically includes the gauge potential and bosonic and fermionic adjoint matter fields. The action is taken to be the integral of a $\uN$ gauge-invariant single-trace local Lagrangian density,
\be\label{backaction}
S_{N}(M) = \frac{N}{\lambda}\int\!\d^{p+1} x\, \tr_{N} L\bigl(M(x),\partial M (x)\bigr)\, .\ee
The subscripts $N$ indicate that the gauge group is $\uN$; in particular, the notation $\tr_{N}$ means that the trace is over $\uN$ indices. We have factorized the 't~Hooft coupling $\la$ in front of the action. In the large $N$ limit, $\la$ is kept constant, together with other couplings that may be included in $L$.

A collection of $N$ D$p$-branes is simply \emph{defined} to be a gauge field theory of the form \eqref{backaction}. The $(p+1)$-dimensional space-time of the gauge theory is the worldvolume of the D$p$-branes. The indices $a,b$ label the different branes and can be thought of as being carried by the endpoints of open strings attached to the D-branes. The fields $M^{a}_{\ b}$ correspond to the excitations of these strings. Since the gauge group is $\uN$, the string theory is oriented. 

Of course, at this stage, the definition of what we mean by D-brane is no more than a choice of terminology. The gauge theory we consider may describe the worldvolume dynamics of a known D-brane configuration in string theory, but this is not required. When no string theory construction exists, it is  still natural for our purposes to use the D-brane language and consider that any gauge theory defines an open string theory describing the dynamics of an abstract collection of D-branes.

The theory we start with may actually be more general than a standard gauge theory like \eqref{backaction}. We could consider an open string field theory, with an infinite number of open string excitations. For example, we could study the string field theory living on a stack of D3 branes in type IIB string theory, without taking the low energy or near horizon limit on the branes. Most of what we are going to say below applies to this more general set-up. The interest in starting with an open string field theory is that, a priori, the dual holographic background is expected to be asymptotically flat.

\subsection{Probe D-branes}
\begin{figure}
\centerline{\includegraphics[width=6in]{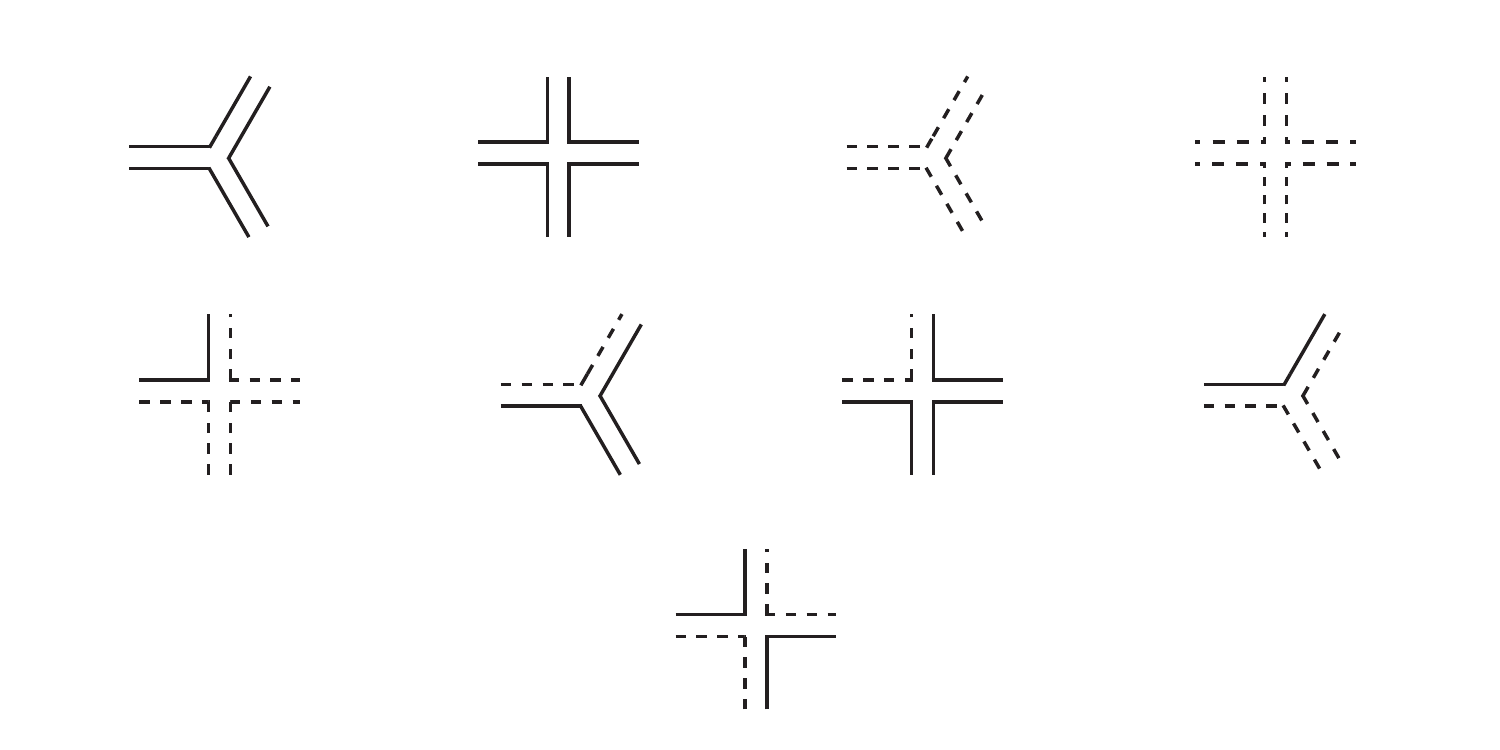}}
\caption{Typical interaction vertices in the background plus probe brane microscopic action \eqref{stot0}. Plain and dashed lines are associated with indices $1\leq a,b,\ldots\leq N$ and $1\leq i,j,\ldots\leq K$ labelling the background branes and the probe branes respectively.
\label{verticesfig}}
\end{figure}

Consider the action \eqref{backaction}, but now for $N+K$ colors instead of $N$, where $N$ and $K$ are a priori arbitrary positive integers. In the D-brane language, this is a model living on a stack of $N+K$ branes. Separate this stack into two sets of $N$ and $K$ branes respectively. Call the $N$ branes in the first stack the ``background'' branes and the $K$ branes in the second stack the ``probe'' branes. This terminology comes in part from the fact that we shall eventually be interested in the limit $N\gg K$. This separation into ``background'' and ``probe'' branes is not a gauge-invariant procedure but, as we have already indicated, we postpone the discussion of this important subtlety to Section \ref{gfsec}. In terms of the $(N+K)\times (N+K)$ Hermitian matrix fields, the distinction between ``background'' and ``probe'' branes amounts to decomposing the matrices according to the $\uN\times\uK\subset\text U(N+K)$ quantum numbers, by writing
\be\label{funddec} M = 
\begin{pmatrix} V^{a}_{\ b} & \bar w^{a}_{\ i}\\ w^{i}_{\ a} &
v^{i}_{\ j}
\end{pmatrix}\, .\ee
The $a,b$ and $i,j$ are $\uN$ and $\uK$ indices respectively. The block-diagonal matrices $V$ and $v$ are Hermitian and transform in the adjoint representations of $\uN$ and $\uK$ respectively. They are associated with open strings attached either to the background or to the probe branes. The off-diagonal block matrices $w$ and $\bar w$ are Hermitian conjugate to each other and transform in bi-fundamental representations. They are associated with open strings for which one endpoint is attached to the background branes and the other endpoint is attached to the probe branes.

By plugging the decomposition \eqref{funddec} into the action $S_{N+K}$, we always find a formula of the form
\be\label{stot0} S_{N+K}(M) = S_{N}(V) +\Bigl(\frac{N}{K}+1\Bigr) S_{K}(v) + S_{N,K}(V,v,w,\bar w)\, .\ee
We have arranged the terms in a way convenient to study the large $N$, fixed $K$ limit later. The ``mixed,'' $\uN\times\uK$ invariant, action $S_{N,K}$ describes the interactions between the background and the probe branes. Typical interaction terms in $S_{N}$ and $S_{K}$ can be written schematically as $\tr V^{3}$, $\tr V^{4}$ and $\tr v^{3}$, $\tr v^{4}$ respectively, whereas $S_{N,K}$ generically includes terms of the form $\bar w v^{2}w$, $w V \bar w$, $w V^{2}\bar w$, $\bar w v w$, $(\bar w w)^{2}$. The corresponding interaction vertices are depicted in Fig.\ \ref{verticesfig}. Note that on these diagrams, the plain and dashed lines, which carry background and probe brane indices $a,b,\ldots$ and $i,j,\ldots$, can also be interpreted as open string worldsheet boundaries belonging to the background and probe branes respectively.

\smallskip 

\noindent\emph{Remarks}

\noindent (i) The notion of ``probe'' branes that we have introduced do not presuppose the existence of a holographic emergent geometry that the ``probe'' D-branes can actually probe. Indeed, a priori, there is no notion of a space ``transverse'' to the background branes and even no elementary adjoint scalar fields in the model.

\noindent (ii) In the context of D-brane configurations in string theory, finding the $\uN\times\uK$ invariant Lagrangian describing the interactions between background and probe branes can be rather non-trivial, involving the computation of complicated disk diagrams with insertions of boundary changing vertex operators. The case discussed above is special, because the background and the probe branes are of the same nature, and thus the full set of $N+K$ branes has a $\text{U}(N+K)$ invariant description given by the action $S_{N+K}$. As we have just explained, the interaction Lagrangian can then be found straightforwardy from field theory by plugging the decomposition \eqref{funddec} into this action. The case of systems containing probe branes and background branes of different nature and dimensionality can also be described purely in gauge theoretic terms, see \ref{GenBsysSec} below. The computation of the interaction Lagrangian is then a non-trivial task both from the string theory and the field theory points of view. 

\subsection{Examples}\label{ex3Sec}

The simplest example one can consider is the zero dimensional Hermitian matrix model with action
\be\label{onemmaction} S_{N}(M) = \frac{N}{\la}\tr_{N} \Bigl(\frac{1}{2}M^{2} + \frac{1}{4}M^{4}\Bigr)\, .\ee
We could generalize this action to $\tr P(M)$, for an arbitrary polynomial $P$, but all the main qualitative features can be discussed with \eqref{onemmaction}. Using the decomposition \eqref{funddec} in the action $S_{N+K}$, we obtain \eqref{stot0} with 
\be\label{Sm1mm}
S_{N,K}(V,v,w,\bar w)  =\frac{N+K}{\la}\Bigl( \bar w w + w V^{2}\bar w + \bar w v^{2} w + \bar w v w V
+\frac{1}{2} \bar w w\bar w w\Bigr)+\frac{K}{N}S_{N}(V)\, .
\ee
The fact that the term $\frac{K}{N}S_{N}$ must be included comes from the particular definition of $S_{N,K}$ given by \eqref{stot0}. It is a convenient convention in the large $N$ limit. In \eqref{Sm1mm}, we use a matrix notation and an overall trace, either over $\uN$ or over $\uK$ indices, is always implicitly assumed. For example, $\bar w w = \bar w^{a}_{\ i} w^{i}_{\ a}$, $w V^{2}\bar w =
w^{i}_{\ a}V^{a}_{\ b}V^{b}_{\ c}\bar w^{c}_{\ i}$, $\bar w w \bar w w = \bar w^{a}_{\ i}w^{i}_{\ b}\bar w^{b}_{\ j}w^{j}_{\ a}$, etc... Note that all the quartic vertices depicted in Fig.\ \ref{verticesfig} are generated in this simple model.

Clearly this procedure can be performed for any matrix gauge theory. Let us work out explicitly the case of the pure Yang-Mills theory, with action
\be\label{pureYMact} S_{N}(\mathscr A) = -\frac{N}{2 \la}\int\!\d^{d} x\, \tr_{N}\mathscr F^{\mu\nu}\mathscr F_{\mu\nu}\ee
and gauge potential $\mathscr A_{\mu}$,
\be\label{Fmunu} \mathscr F_{\mu\nu} = \partial_{\mu}\mathscr A_{\nu}
-\partial_{\nu}\mathscr A_{\mu} + i[\mathscr A_{\mu},\mathscr A_{\nu}]\, .\ee
As above, we start with the $\text{U}(N+K)$ model and decompose
\be\label{AYMdec} \mathscr A_{\mu} = 
\begin{pmatrix} \mathscr V^{\, a}_{\mu\, b} & \frac{1}{\sqrt 2} \bar W^{\, a}_{\mu\, i}\\ 
\frac{1}{\sqrt{2}} W^{\, i}_{\mu\, a} & V^{\, i}_{\mu\, j}\end{pmatrix}\, ,\ee
with $\bar W_{\mu} = W_{\mu}^{\dagger}$.
Plugging \eqref{AYMdec} into \eqref{pureYMact} then yields an action of the form \eqref{stot0}, with
\begin{multline}\label{SmYM}S_{N,K}(\mathscr V,V,W,\bar W)  =\\\frac{N+K}{\la} \int\!\d^{d}x\, \biggl[
-\nabla^{\mu}\bar W^{\nu}\bigl(\nabla_{\mu}W_{\nu}-\nabla_{\nu}W_{\mu}\bigr) +i W^{\mu}G_{\mu\nu}\bar W^{\nu}+ i \bar W^{\mu}F_{\mu\nu}W^{\nu}\\
+ \frac{1}{4}\Bigl( 2\bar W^{\mu}W^{\nu}\bar W_{\mu}W_{\nu}
-\bar W^{\mu}W_{\mu}\bar W^{\nu}W_{\nu} - \bar W^{\mu}W^{\nu}\bar W_{\nu}W_{\mu}\Bigl)\biggr]+\frac{K}{N}S_{N}(\mathscr V)\, .
\end{multline}
As in \eqref{Sm1mm}, an overall trace is always assumed in \eqref{SmYM}. The fields $G_{\mu\nu}$ and $F_{\mu\nu}$ are the $\uN$ and $\uK$ field strengths respectively,
\begin{align}\label{Gdef} G_{\mu\nu} = \partial_{\mu}\mathscr V_{\nu}- 
\partial_{\nu}\mathscr V_{\mu} + i[\mathscr V_{\mu},\mathscr V_{\nu}]\, ,\\
\label{Fdef} F_{\mu\nu} = \partial_{\mu} V_{\nu}- 
\partial_{\nu} V_{\mu} + i[ V_{\mu}, V_{\nu}]\, .
\end{align}
The $\uN\times\uK$ covariant derivative acts, in matrix notation, as
\be\label{nablaYMdef}\nabla_{\mu}W_{\nu} = \partial_{\mu}W_{\nu} + i V_{\mu} W_{\nu} - i W_{\nu}\mathscr V_{\mu}\, ,\quad 
\nabla_{\mu}\bar W_{\nu} = \partial_{\mu}\bar W_{\nu} + i \mathscr V_{\mu} \bar W_{\nu} - i \bar W_{\nu}V_{\mu}\, .\ee
Before gauge fixing, Eq.\ \eqref{SmYM} defines the interactions between background and probe Yang-Mills D-branes. Note that all the vertices depicted in Fig.\ \ref{verticesfig} are generated.

\subsection{The dual description\label{dualdesSec}}

So far, the ``background'' and ``probe'' branes have played a strictly parallel 
r\^ole. The distinction between the two comes from the condition 
\be\label{backvsprobe} N\gg K\, .\ee
In practice, we pick a fixed number $K$ of probe branes and let $N\rightarrow\infty$, keeping the 't~Hooft coupling $\la$ fixed. In this limit, the intuition, which is supported by the AdS/CFT framework, is that there should exist a dual and more appropriate description of the background plus probe brane system which replaces the microscopic definition given by the action \eqref{stot0}. In this dual description, the background branes should disappear altogether. In particular, no variable can carry the indices $a,b$, etc., anymore. The background branes are replaced by a non-trivial closed string classical background, with new classical emerging dimensions of space. The probe branes can literally probe this background, which justifies a posteriori the terminology.

 The effective action for the probe branes should thus match the non-abelian effective action for D-branes in a non-trivial closed string background, see e.g.\ \cite{Myers}. In particular, this action must include adjoint scalar fields associated with the emergent target space matrix coordinates. More accurately, the full description of the probe branes should correspond to a full-fledged open string field theory, with an infinite tower of excited string states, matching the spectrum of the open strings attached to the probe branes embedded in the emergent closed string dual geometry. 

These claims, in the context of standard superstring theory, essentially follow from the usual AdS/CFT lore in which probe branes are added \cite{fer1}. They have been tested in details on a handful of non-trivial examples \cite{fer1,fer3,fer4,fer5}, allowing to accurately derive full supergravity backgrounds from the microscopic computations of the probe brane actions. One of the most important point we are making in the present paper is that the same conceptual framework can be consistently considered for a very general notion of D-brane, much more general that the one usually dealt with in string theory, encompassing any gauge theory. This is a crucial and non-trivial step which, if valid, allows to convey many ideas from the usual D-brane/holographic physics in string theory to more realistic and interesting field theories, greatly extending the field of application of these ideas and potentially making connections with other approaches. 

Of course, at this stage, it is not yet clear how the holographic description of the probe branes can emerge from the field theory action \eqref{stot0}. Sketchily, the probe brane action $\mathscr A_{N,K}$ should be 
obtained by integrating out the background and mixed variables,
\be\label{sefftentative} e^{-\mathscr A_{N,K}}\overset{?}{=} 
\frac{e^{-(\frac{N}{K}+1)S_{K}(v)}}{Z_{N}}\int\!\mathcal D V\mathcal D w\mathcal D \bar w\, e^{-S_{N}(V)-S_{N,K}(V,v,w,\bar w)}\, ,\ee
where 
\be\label{ZNdef1} Z_{N}= \int\!\mathcal D V\, e^{-S_{N}(V)}\ee
is the partition function of the gauge theory. With this definition, $\mathscr A_{K,N}$ depends only on $v$. In the familiar supersymmetric cases, like the $\nn=4$ gauge theory, the set of fields $v$ contains scalars and we automatically get an emerging space on which the probe branes can move. However, when the set of fields $v$ do not contain scalars, for example for the pure Yang-Mills theory, Eq.\ \eqref{sefftentative} cannot be consistent with the emerging space picture. In which sense is it then natural or useful to include scalar fields in the effective description of the probe branes, in the limit \eqref{backvsprobe}? Where do these degrees of freedom, together with the infinite tower of open string states that live on the probe branes, come from? We shall provide simple answers to these questions in Section \ref{diagsec}.

\subsection{More general background plus probe D-brane systems\protect\footnote{This subsection lies somewhat out of the paper's main line of development and may be omitted in a first reading.}\label{GenBsysSec}}

Our main emphasis is on background plus probe D-brane systems for which the background and the probe branes are parallel and of the same nature. Such systems are in some sense canonical, since they can always be built starting from any gauge theory. They also have special properties that make them particularly interesting, for example the relation \eqref{fundrel}, which we shall derive later, and which shows that the probe effective action encodes all the information about the gauge theory. However, much more general brane configurations are of course common in string theory. In particular, background plus probe D-brane systems for which the background and the probe branes have different dimensionality can also be very useful to study holography \cite{fer1,fer3,fer4,fer5}, following exactly the same philosophy as the one reviewed in the previous subsection. It is thus interesting to describe this more general notion of D-brane systems in purely field theoretic terms as well.

A system of $N$ background D$p$-branes, on which the gauge theory we start with live, together with $K$ probe D$q$-branes, $q\leq p$, can be \emph{defined} as any gauge-theoretic set-up whose dynamics is governed by a microscopic action which is the sum
\be\label{STot} S_{N} + \tilde S_{K} + S_{N,K}\ee
of three terms having the following properties. One term is the background brane action, which is the same as the gauge theory action  \eqref{backaction} we study. Another term describes the probe branes dynamics, which must be governed by an action $\tilde S_{K}(\tilde M)$ of the form \eqref{backaction}, but a priori with a different worldvolume dimension $q\leq p$ and Lagrangian $\tilde L$,
\be\label{probeaction} \tilde S_{K}(\tilde M)=\frac{K}{\tilde\la}\int\!\d^{q+1}x\,\tr_{K}\tilde L\bigl( \tilde M(x),\partial \tilde M(x)\bigr)\, .\ee
The worldvolume of the probe branes is naturally a subset of the space-time of the original gauge theory, i.e.\ a subset of the background branes worldvolume.
Finally, the third term $S_{N,K}$ describes the interactions between the background and the probe branes,
\be\label{SNKgen} S_{N,K}(M,\tilde M, w,\bar w)=\int\!\d^{q+1}x\, L_{N,K}\bigl(M,\tilde M,w,\bar w,\partial M,\partial\tilde M,\partial w,\partial\bar w\bigr)\, .\ee
This term depends on $\uN\times\uK$ bifundamental fields $w$ and $\bar w$, which are associated with the strings stretched between the background and the probe branes and which live at the intersection of the background and probe worldvolumes. Typically in a standard gauge theoretic construction this intersection is the same as the probe brane worldvolume and it coincides with the domain of integration in \eqref{probeaction}. Note that the diagrams depicted in Fig.\ \ref{verticesfig} still yield the typical interaction vertices in the more general background plus probe D-brane system just described.

The above point of view may seem abstract, but it is actually easy to come up with gauge theory constructions of precisely this type. For example, let us consider instantons in a four dimensional gauge theory. We can define a system of background D3-branes with probe D(-1)-branes (or D-instantons). The background brane fields $M^{a}_{\ b}$ automatically include the gauge potential 
$A_{\mu}$, whereas the probe brane fields $\tilde M^{i}_{\ j}$ (which, in the case of instantons, could be more properly called moduli) always include the instantons center-of-mass matrix coordinates $X_{\mu}$. The mixed fields $w^{i}_{\ a}$ and $\bar w^{a}_{\ i}$ correspond to bosonic spinor variables $q_{\alpha}$ and $\bar q^{\alpha}$ which, together with $X_{\mu}$, form the minimal set of ADHM moduli. When more fields are present on the background branes, then the instantons can also have more moduli. For example, in the case of the $\nn=4$ gauge theory, one must include supersymmetric partners and the full set of variables read
\be\label{N4ivaviables}
\begin{split}
\bigl\{M^{a}_{\ b}\bigr\} & = \bigl\{A_{\mu}, \lambda_{\alpha A},\bar\lambda^{\dot\alpha A},\varphi_{m}\bigr\}\, ,\\
\bigl\{\tilde M^{i}_{\ j}\bigr\} &= \bigl\{ X_{\mu}, \psi^{\dot\alpha A}\bigr\}\, , \\
\bigl\{w^{i}_{\ a},\bar w^{a}_{\ i}\bigr\} &= \bigl\{
q_{\alpha},\chi^{A},\bar q^{\alpha},\bar\chi^{A}\bigr\}\, ,
\end{split}\ee
where $\lambda$, $\bar\lambda$, $\psi$, $\chi$ and $\bar\chi$ are fermionic, $A$ a $\text{Spin}(6)$ spinor index, or equivalently an $\text{SU}(4)$ fundamental (upper) or antifundamental (lower) index, and $m$ an $\text{SO}(6)$ vector index. Furthermore, the ADHM moduli satisfy some constraints. These constraints may be implemented by introducing Lagrange multipliers, which are variables of the form $\tilde M^{i}_{\ j}$ living on the probe branes. For example, the standard bosonic ADHM constraint is associated with a self-dual bosonic $D_{\mu\nu}$, whereas its supersymmetric partner in the $\nn=4$ theory is a fermionic $\La_{\alpha A}$. 

The strategy to derive the action \eqref{SNKgen} for the D3/D-instanton system is, at least in principle, straightforward. The background action $S_{N}$ is of course the action $S_{\text{YM}}$ of the Yang-Mills theory we consider, for example the pure Yang-Mills theory or the $\nn=4$ model. If we note
\be\label{instsol} M = \mathscr M (\tilde M,w,\bar w)\ee
the most general instanton solution for the Yang-Mills fields $M$ as a function of the moduli $\tilde M,w,\bar w$, the full action \eqref{STot} for the background plus brane system is then given by
\be\label{Sinst1} S\bigl(M,\tilde M,w,\bar w\bigr) = S_{\text{YM}}\bigl(M + \mathscr M(\tilde M,w,\bar w)\bigr)\, .\ee
One must also determine the integration measure over the moduli $\tilde M,w,\bar w$ from the gauge theory integration measure $\mathcal D M$. In the $\nn=4$ theory, this integration measure is flat.

Of course, even though the above procedure is conceptually simple, its detailed technical implementation can be rather complicated. For example, to my knowledge, it has never been carried through in the case of the $\nn=4$ gauge theory, for arbitrary $\nn=4$ background fields $\{M\}=\{A_{\mu}, \lambda_{\alpha A},\bar\lambda^{\dot\alpha A},\varphi_{m}\}$. In this case, the background action $S_{N}(M)=S_{\nn=4}$ is the $\nn=4$ super Yang-Mills action and the sum of the probe action plus the interacting piece $S_{N,K}$  at zero background fields $M=0$ can be written as (see e.g.\ \cite{instrev} and references therein)
\be\label{probemixedinst}  -\frac{1}{2}\tr_{\text U(K)}
v_{m}O^{-1}v_{m}\, ,\ee
where
\be\label{vmform} v_{m} = \frac{1}{2}\chi^{A}\bar\chi^{B}\Sigma_{mAB} +\frac{4\pi^{2}}{g^{2}}\bigl[\psi^{\dot\alpha A},\psi_{\dot\alpha}^{\ B}\bigr]\Sigma_{mAB}\ee
and $O$ is a moduli-dependent linear operator acting on $K\times K$ matrices $m$ as
\be\label{Olindef} O\cdot m = \frac{8\pi^{2}}{g^{2}}\bigl[X_{\mu},[X_{\mu},m]\bigr] + \frac{1}{2}\bigl\{q_{\alpha}\bar q^{\alpha},m\bigr\}\, .\ee
The matrices $\Sigma_{mAB}$ are six dimensional versions of the Pauli matrices (explicit definitions can be found e.g.\ in \cite{fer1}), the bracket in \eqref{vmform} is a graded commutator and $g$ is the Yang-Mills coupling constant. Together with the ADHM constraints, \eqref{probemixedinst} generate interaction vertices of the form depicted in Fig.\ \ref{verticesfig}. If, moreover, the background $\nn=4$ fields are turned on, more vertices must be included, for example
\be\label{instgenvertices} \bar q\sigma_{\mu\nu}qF_{\mu\nu}\, ,\
q\varphi_{m}\varphi_{m}\bar q\, ,\ (q\varphi_{m}\bar q)^{2}\, ,\
\bar\chi\Sigma_{m}\chi\varphi_{m}\, ,\ \chi\lambda\bar q\, ,\ q\lambda\bar\chi\, ,\ (\bar\chi\Sigma_{m}\chi)^{2}\, ,\ \chi\Sigma_{m}\bar\chi q\varphi_{m}\bar q\, ,\text{etc...}\ee
In these formulas, $\sigma_{\mu\nu}$ represents the generators of self-dual Euclidean space-time rotations. Moreover, the $\nn=4$ fields $F_{\mu\nu}$, $\lambda$ and $\varphi$ are evaluated at the location of the instantons, which is given by $\frac{1}{K}\tr X_{\mu}$ in terms of the moduli $X_{\mu}$. The precise form of these couplings have been determined by string theory methods is \cite{couplings}. See e.g.\ \cite{fer1} and references therein for more details.

One can repeat the above discussion for objects in field theory different from instantons, for example solitons or domain walls. \emph{Each time an action of the form \eqref{STot} is obtained, the model can be interpreted as a background plus probe D-brane system.} This is the case, for instance, for 't~Hooft-Polyakov monopoles in four dimensional Yang-Mills models with adjoint Higgs fields. The background variables $M$ then correspond as usual to the elementary fields in the Yang-Mills-Higgs model, the probe variables $\tilde M$ to the Nahm's data and the mixed variable $w$ and $\bar w$ to quantum mechanical degrees of freedom living on the monopole worldlines. In this set-up, the Higgs vaccum expectation value can be interpreted as the transverse separation between stacks of D3-branes and the monopoles are stretched D-strings between these stacks.  A detailed discussion of this system can be found in \cite{diac}. The full derivation of the corresponding action \eqref{STot} from first principles is an interesting open problem.

\section{Diagrammatics}\label{diagsec}
\begin{figure}
\centerline{\includegraphics[width=6in]{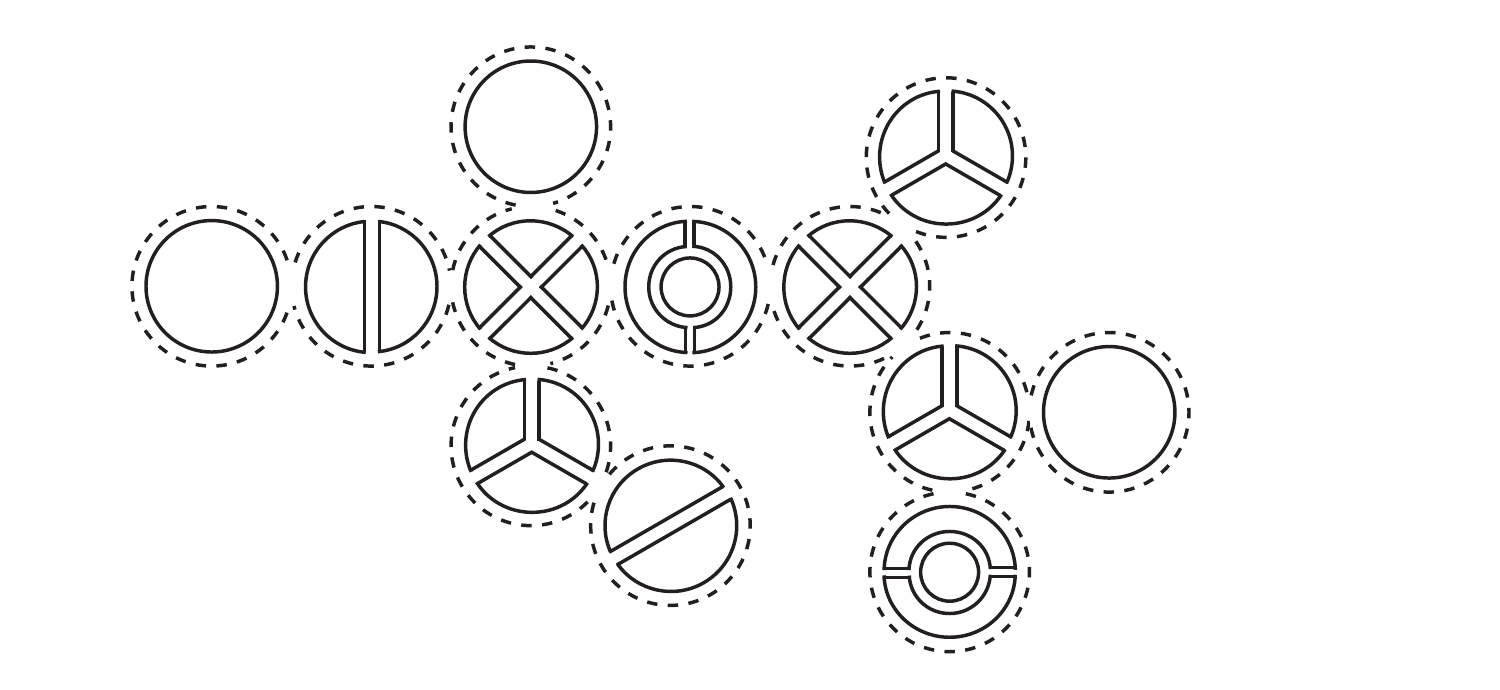}}
\caption{A typical leading Feynman diagram, of order $O(N)$, contributing to the large $N$ probe branes effective action in 't~Hooft's double-line representation. We use vacuum-like diagrams, the dependence on the probe brane fields being included in the propagators for the mixed variables $w,\bar w$ (see Eqs.\ \eqref{b0mm} and \eqref{rhorop} for an example). This is equivalent to considering diagrams with arbitrary insertions of $\text{U}(K)$ adjoint probe brane fields. 
\label{typicalDfig}}
\end{figure}
\subsection{Trees of dressed bubbles}\label{treebubbleSec}

Let us put the vertices depicted in Fig.\ \ref{verticesfig} together to build Feynman diagrams for the probe D-branes effective action. A typical diagram that contributes to the leading large $N$ approximation is depicted in Fig.\ \ref{typicalDfig}, using 't~Hooft's double-line representation. 

\begin{figure}
\centerline{\includegraphics[width=6in]{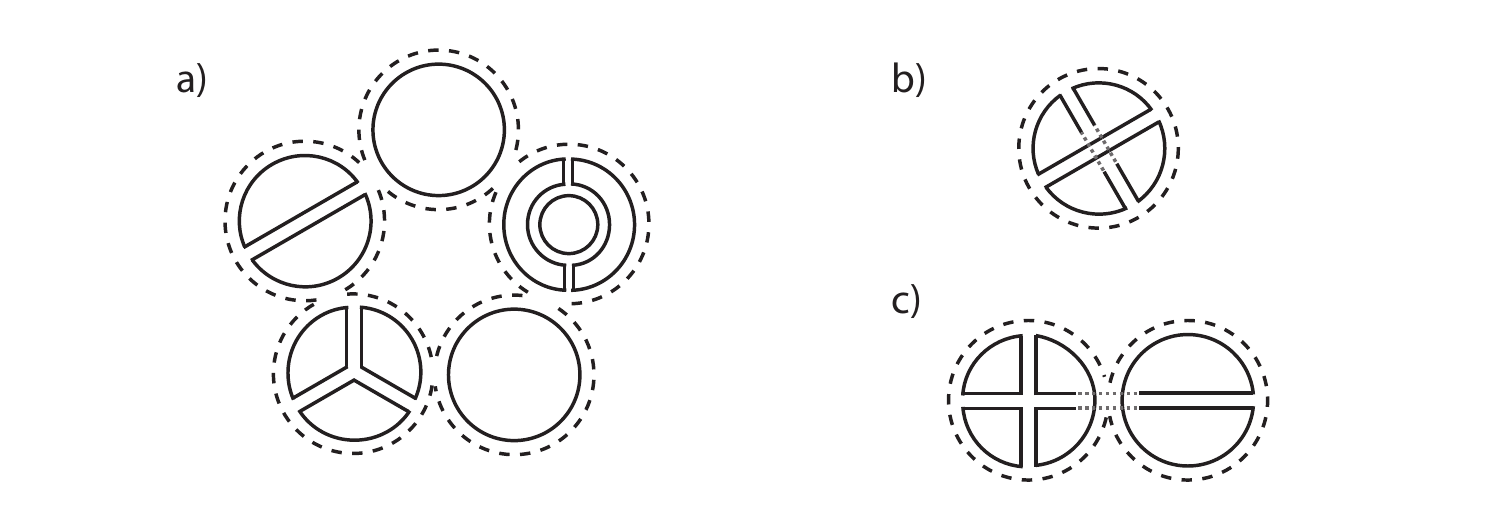}}
\caption{Subleading diagrams: a) a loop of bubbles, contributing at order $O(1)$; b) a non-planar contribution to the bubble, of order $O(1/N)$; c) a non-planar contribution linking two different bubbles, of order $O(1/N)$.
\label{subleadfig}}
\end{figure}

The large $N$ counting for the effective action is very similar to the standard large $N$ counting for vacuum diagrams in an ordinary matrix model. The difference is that, in our case, we have two different types of indices corresponding to plain and dashed lines, associated with the background and probe branes respectively. Each closed plain line loop yields a factor of $N$ whereas each closed dashed line loop yields a factor of $K$. When $N\gg K$, the dashed line loops are thus subleading and the large $N$ limit is dominated by diagrams containing only one such loop (diagrams with no dashed line loop at all can also contribute to the probe brane effective action at order $N$ and lower, because of terms like the last term in \eqref{Sm1mm} or \eqref{SmYM}; they yield only constant, field-independent, contributions). In the leading large $N$ limit, we thus keep the diagrams which are planar \emph{and} contain only one closed loop made of dashed lines. All these diagrams are proportional to $N$, as expected for a D-brane effective action. Examples of subleading contributions are depicted in Fig.\ \ref{subleadfig}. To all orders in $1/N$, one finds an expansion of the precise form \eqref{AKNexp}.

The structure of the leading Feynman diagrams found above is particularly interesting. They are made of ``dressed bubbles'' which are themselves put together in a tree-like shape. The dressing of the bubbles comes from the matrix model structure and is associated with the path integral over the background brane matrix fields $V^{a}_{\ b}$. The trees of bubbles are associated with the path integral over the mixed variables $w^{i}_{\ a}$ and $\bar w^{a}_{\ i}$, which behave as large $N$ vector model fields when $N\gg K$. The overall structure reflects the mixed matrix model/vector model nature of the background plus probe D-brane action \eqref{stot0}.

\begin{figure}
\centerline{\includegraphics[width=6in]{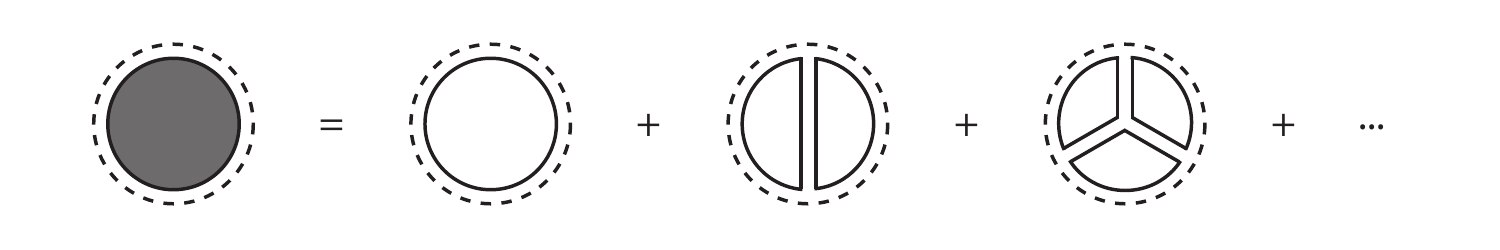}}
\caption{A fully dressed bubble is given by a purely one-loop contribution (the ``bare'' bubble) plus an infinite series of planar diagrams generated by the couplings between the mixed variables $w,\bar w$ and the fields living on the background branes.
\label{dressedbubble}}
\end{figure}

\begin{figure}
\centerline{\includegraphics[width=6in]{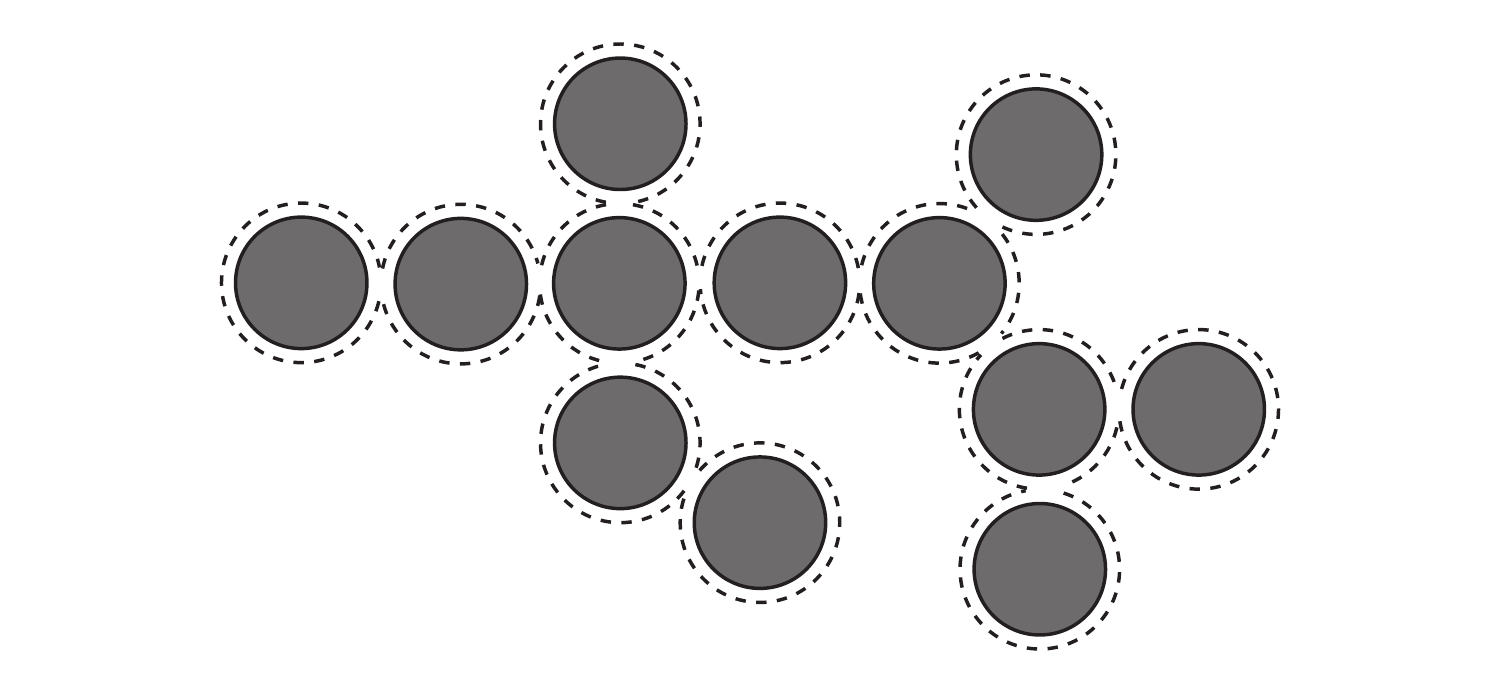}}
\caption{A tree of dressed bubbles, corresponding to the sum of all the diagrams having the same tree shape as the diagram in Fig. \ref{typicalDfig}, but with all possible dressings in the bubbles included. The probe brane effective action is given by a sum over such trees.
\label{treefig}}
\end{figure}

One can actually always define a notion of $p$-valent dressed bubble $b_{p}(v)$, corresponding to a bubble that can be attached to $p$ other bubbles and for which the full dressing is taken into account, see Fig.\ \ref{dressedbubble}. Explicit formulas for the one matrix model are provided below. The sum over all possible diagrams of the type depicted in Fig.\ \ref{typicalDfig}, which includes all the possible dressings in the bubbles and all the possible trees of bubbles, can then be rewritten as a sum over trees of dressed bubbles, as in Fig.\ \ref{treefig}. Such a simple picture is of course only valid at leading order $O(N)$. For example, at order $O(1)$, one must include diagrams containing one loop of bubbles, as e.g.\ the diagram a in Fig.\ \ref{subleadfig}; at order $O(1/N)$, one must take into account not only diagrams with two loops of bubbles and non-planar corrections to the dressing of the bubbles (as in the diagram b in Fig.\ \ref{subleadfig}), but also links between bubbles, as in the diagram c of the same figure.


Let us illustrate the structure just explained on the simple example of the one matrix model. We thus consider the action \eqref{stot0} with $S_{N}$ and $S_{N,K}$ given by \eqref{onemmaction} and \eqref{Sm1mm}. The dressed bubble of valence zero is obtained by integrating out $w$ and $\bar w$ without taking into account the quartic coupling $(\bar w w)^{2}$ and then integrating over $V$,
\be\label{b0mm} b_{0}(v) = \frac{1}{N}\ln\Bigl\langle\det\bigl( \mathbb I_{KN} +v^{2}\otimes\mathbb I_{N}+v\otimes V+\mathbb I_{K}\otimes V^{2}\bigr)\Bigr\rangle\, . \ee 
We use the notation $\mathbb I_{r}$ for the unit $r\times r$ matrix and we denote by $\langle\cdot \rangle$ expectation values in the background brane theory, which here is simply the matrix model with action $S_{N}(V)$. Similarly, the dressed bubble of valence $p$ is found by noting that it contains $p$ propagators of the $w,\bar w$ fields joining the $p$ vertices on the bubble. The propagator is given by
\be\label{rhorop} \rho (v,V) = \bigl(\mathbb I_{KN} +v^{2}\otimes\mathbb I_{N}+v\otimes V+\mathbb I_{K}\otimes V^{2}\bigr)^{-1}\ee
and thus $b_{p}(v)$, which is a cyclic tensor, can be written as
\be\label{bpmm} b_{p}(v)^{i_{1}\cdots i_{p}}_{\ j_{1}\cdots j_{p}} = 
\frac{1}{N}
\Bigl\langle\tr_{N} \rho^{i_{1}}_{\ j_{1}}\cdots \rho^{i_{p}}_{\ j_{p}}
\Bigr\rangle\, ,\ee
where matrix multiplications and the trace in \eqref{bpmm} are over $\uN$ indices only. The trees are then obtained by gluing together the bubbles $b_{p}$. For example, the tree depicted in Fig.\ \ref{treefig} is given by
\begin{multline}\label{treegluing} (b_{1})^{i_{1}}_{\ i_{2}}(b_{2})^{i_{2}k_{4}}_{\ i_{3}i_{1}}(b_{4})^{i_{3}i_{7}k_{1}k_{3}}_{\ i_{4}i_{8}k_{2}k_{4}}
(b_{2})^{i_{4}i_{6}}_{\ i_{5}i_{7}} (b_{1})^{i_{5}}_{\ i_{6}}
(b_{1})^{k_{2}}_{\ k_{3}} (b_{2})^{i_{8}j_{9}}_{\ i_{9}k_{1}}\\
(b_{3})^{i_{9}j_{6}j_{8}}_{\ j_{1}j_{7}j_{9}} 
(b_{3})^{j_{1}j_{3}j_{5}}_{\ j_{2}j_{4}j_{6}}
(b_{1})^{j_{2}}_{\ j_{3}} (b_{1})^{j_{4}}_{\ j_{5}} (b_{1})^{j_{7}}_{\ j_{8}}\, .
\end{multline}

\subsection{Summing the trees and emergent dimensions}\label{treesumsec}

Let us assume, for the moment, that the dressed bubbles can be computed, either exactly or within some approximation. There remains to perform the sum over all the possible trees of bubbles. Maybe not surprisingly, but most interestingly, \emph{this sum can always be done exactly.} 

\begin{figure}
\centerline{\includegraphics[width=6in]{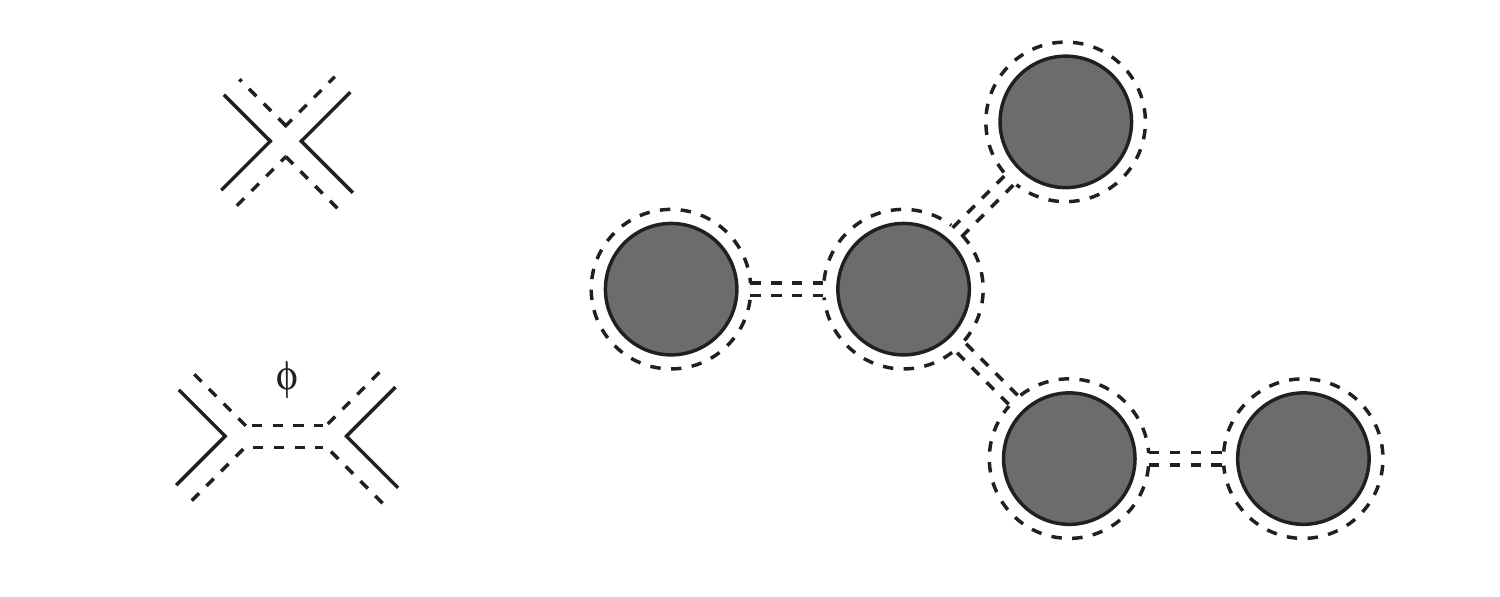}}
\caption{By introducing new degrees of freedom $\Phi$ in the adjoint of $\uK$, the interaction vertices involving the mixed variables can be replaced by new vertices in which these variables appear only quadratically (left inset). The trees of bubbles then become genuine tree Feynman graphs in a non-local field theory which includes $\Phi$ and whose interaction vertices are the dressed bubbles (right inset).\label{realtreefig}}
\end{figure}

The basic trick is well known from the study of the vector models (see e.g.\ \cite{vectorrev, Neumann} for reviews and examples). Remember that, in our case, the vector-like variables are the $w^{i}_{\ a}$ and $\bar w^{a}_{\ i}$, the $\uN$ index $a$ playing the r\^ole of the vector index. One then introduces new degrees of freedom which are $\uN$ invariant bilinears,
\be\label{phigenform}\Phi^{i}_{\ j}\sim w^{i}_{\ a}\bar w^{a}_{\ j}\, .\ee
Note that these new variables are automatically in the adjoint of $\uK$. Introducing the $\Phi^{i}_{\ j}$ allows to replace the interaction vertices involving the mixed variables $w$, $\bar w$ by new vertices in which these variables appear only quadratically. The Gaussian integration over the mixed variables can then be performed exactly. The resulting model, which automatically includes the new variables $\Phi$ in the adjoint of $\uK$, has (non-local) interaction vertices given by the dressed bubbles. In the large $N$ limit, this description becomes classical and the corresponding tree Feynman graphs of course coincide with the trees of bubbles of the original model. This mechanism is depicted in Fig.\ \ref{realtreefig}. In particular, the quantum fluctuations of the bilinears \eqref{phigenform} are suppressed when $N\rightarrow\infty$, a property which can also be understood by noting that they are sums of a large number of random variables.

From the path integral point of view, Eq.\ \eqref{sefftentative}, the new variables $\Phi$ are introduced in such a way that
\be\label{pathin1} e^{-S_{N,K}(V,v,w,\bar w)} = \int\!\mathcal D\Phi\,
e^{-\hat S_{N,K}(V,v,\Phi,w,\bar w)}\, ,\ee
where the action $\hat S_{N,K}$ is \emph{quadratic} in the mixed fields $w,\bar w$. The probe D-brane effective action can then be defined by the equation
\be\label{seffbetter} e^{-\mathscr A_{N,K} (v,\Phi)}=\frac{e^{-(\frac{N}{K}+1)S_{K}(v)}}{Z_{N}}\int\!\mathcal D V\mathcal D w\mathcal D\bar w\, e^{-S_{N}(V)-\hat S_{N,K}(V,v,\Phi,w,\bar w)}\, .\ee
This equation replaces the tentative definition \eqref{sefftentative}. The definition of the effective action will be further refined in Section \ref{gfsec}, to take into account gauge invariance, but this will not change the basic features that we are discussing here.

By construction, the integral over $w,\bar w$ in \eqref{seffbetter} is Gaussian and yields a $\uK\times\uN$ invariant functional determinant $\Delta$ depending on $V$, $v$ and $\Phi$,
\be\label{pathint2} \Delta (V,v,\Phi)= \int\!
\mathcal D w\mathcal D\bar w\, e^{-\hat S_{N,K}(V,v,\Phi,w,\bar w)}\, .\ee
Being the result of integrating over $\sim N$ random Gaussian variables, $\ln\Delta$ is automatically proportional to $N$. The definition \eqref{seffbetter} is then equivalent to
\be\label{seffbetter2} \mathscr A_{N,K} (v,\Phi) =\Bigl(\frac{N}{K}+1\Bigr) S_{K}(v) -\ln\bigl\langle\Delta (V,v,\Phi)\bigr\rangle\, ,\ee
where $\langle\Delta\rangle$ denotes the vacuum expectation value of the $\uN$ gauge invariant operator $\Delta$ in the background brane theory. 

In the leading large $N$ limit, the formula \eqref{seffbetter2} further simplifies. The familiar large $N$ factorization property of the  expectation value of product of gauge invariant operators indeed implies that $\ln\langle\Delta\rangle = \langle\ln\Delta\rangle$ in this limit. Using \eqref{leadinglargeN}, Eq.\ \eqref{seffbetter2} thus yields
\be\label{seffbetter3} A_{K} (v,\Phi) = \frac{1}{K} S_{K}(v) -\frac{1}{N}\bigl\langle\ln\Delta(V,v,\Phi)\bigr\rangle_{N=\infty}\, .\ee
The effective action $\mathscr A_{N,K}\simeq NA_{K}$ so obtained is always proportional to $N$. It can be treated classically in the large $N$ limit.

\noindent\emph{Lesson}: we have shown that, on top of the original variables $v$, the probe D-brane effective action naturally depends on additional adjoint variables $\Phi\sim w\bar w$. We propose that \emph{the scalars amongst these new variables are natural candidates to describe emergent space matrix coordinates.} This is a natural mechanism to generate a holographic dual, \emph{even in models that do not contain elementary scalar fields.}

\noindent Before discussing this proposal further, let us illustrate the formalism on a few examples.

\subsection{Examples}\label{exsec4sec}

\subsubsection*{The one matrix model}

The simplest way to make the matrix model action \eqref{Sm1mm} quadratic in $w,\bar w$ is to rewrite it as
\be\label{Sm1mmtilde} \hat S_{N,K} =\frac{N+K}{\la}\Bigl(-\frac{1}{2}\tr_{K}\phi^{2}+ \bar w w +wV^{2}\bar w + \bar w
(v^{2}+\phi)w +\bar w vwV\Bigr) + \frac{K}{N}S_{N}(V)\, .\ee
Indeed, integrating out the $\uK$ adjoint variable $\phi$ correctly yields \eqref{Sm1mm}, with the identification
\be\label{phidef1mm} \phi = w\bar w\, .\ee
Performing the integrals in \eqref{seffbetter} and working to leading order at large $N$, we get
\begin{multline}\label{Seffmm1} A_{K} (v,\phi) = \frac{1}{K}S_{K}(v) +
\frac{K}{N^{2}}\bigl\langle S_{N}(V)\bigr\rangle-\frac{1}{2\la}\tr_{K}\phi^{2}
\\+\frac{1}{N} \Bigl\langle\tr\ln\bigl(\mathbb I_{KN} + (v^{2}+\phi)\otimes\mathbb I_{N} + v\otimes V + \mathbb I_{K}\otimes V^{2}\bigr)\Bigr\rangle\, .
\end{multline}
Using the definitions of the dressed bubbles in \eqref{b0mm} and \eqref{bpmm}, using the large $N$ factorization as well, we can rewrite this action as
\begin{multline}\label{Seffmm2} A_{K} (v,\phi) = \frac{1}{K}S_{K}(v)+\frac{K}{N^{2}}\bigl\langle S_{N}(V)\bigr\rangle -\frac{1}{2\la}\tr_{K}\phi^{2}\\ +b_{0}(v)- \sum_{p\geq 1}\frac{(-1)^{p}}{p} b_{p}(v)^{i_{1}\cdots i_{p}}_{\ j_{1}\cdots j_{p}}\phi^{j_{1}}_{\ i_{2}}\phi^{j_{2}}_{\ i_{3}}\cdots\phi^{j_{p-1}}_{\ i_{p}}\phi^{j_{p}}_{\ i_{1}}\, .
\end{multline}
The tree graphs of the probe action we find are precisely the trees of bubbles, as depicted on the right-hand side of Fig.\ \ref{realtreefig}, with vertices given by the dressed bubbles $b_{p}$. This is perfectly in line with the general discussion. Summing over all the trees of bubbles then simply amounts to solving the classical equation of motion for $\phi$,
\be\label{classeq1} \frac{\partial A_{K}}{\partial\phi} = 0\, .\ee

There is one apparent drawback with the simple procedure just outlined: the field $\phi$ is unstable because of the minus sign in front of the $\tr\phi^{2}$ term in \eqref{Seffmm2}. In some cases, this instability may be waived by the quantum corrections, but this doesn't occur for the one matrix model. It is important to realize that this is not an inconsistency. It simply means that the path integral over $\phi$ must be made over the imaginary axis in $\phi$-space. This subtlety disappears altogether on-shell. The saddle point equation \eqref{classeq1} indeed yields a Hermitian $\phi$ as a function of $v$. However, if one wants to keep $\phi$ off-shell,  it may be desirable to have a stable potential.

There is a simple method that allows to achieve this, by enforcing the relation \eqref{phidef1mm} off-shell with the help of a Lagrange multiplier $L$. We thus modify the prescription \eqref{Sm1mmtilde} by writing
\be\label{Smhat1mm} \hat S_{N,K} =\frac{N+K}{\la}\biggl[ \bar w w + w V^{2}\bar w + \bar w A^{2}w + \bar w A w V + \tr_{K}\Bigl(\frac{1}{2}\phi^{2} - L\bigl(\phi - w\bar w\bigr)\Bigr)\biggr]+\frac{K}{N}S_{N}(V)\, .
\ee
The term $+\frac{1}{2}\tr_{K}\phi^{2}$ automatically reproduces the quartic coupling $\frac{1}{2}\bar w w\bar w w$. To leading order at large $N$, we get in this way
\begin{multline}\label{Seffhat1mm}
A_{K} (v,\phi,L) = \frac{1}{K}S_{K}(v) +
\frac{K}{N^{2}}\bigl\langle S_{N}(V)\bigr\rangle + \frac{1}{\la}\tr_{K}\Bigl(\frac{1}{2}\phi^{2}-L\phi\Bigr)
\\+ \frac{1}{N}\Bigl\langle\tr\ln\bigl(\mathbb I_{KN} + (v^{2}+L)\otimes\mathbb I_{N} + v\otimes V + \mathbb I_{K}\otimes V^{2}\bigr)\Bigr\rangle\, .\end{multline}
Integrating out the Lagrange multiplier by solving 
\be\label{saddleforL} \frac{\partial A_{K}}{\partial L} = 0\, ,\ee
we get a stable potential as a function of $\phi$. 
We shall see in Sec.\ \ref{sfsec} that this procedure has a simple interpretation and can be easily generalized.

Note that if we integrate out $\phi$ from \eqref{Seffhat1mm}, we get $\phi=L$ and we are back to the ``old'' effective action \eqref{Seffmm1} with the unstable potential. The new prescription is to integrate out $L$ instead, which yields a stable action. Of course, on-shell, all these procedures are strictly equivalent. For the one-matrix model, the most natural object to consider is actually the effective action $A_{K}(v)$ as a function of $v$ only, obtained by solving both \eqref{saddleforL} and 
\be\label{saddleforphi}\frac{\partial A_{K}}{\partial\phi} = 0\, .\ee
When $K=1$, the traditional (albeit na\"\i ve in general, see below) interpretation of the effective action $A_{1}=A(v)$ is as the potential seen by one eigenvalue of the matrix $M$ in \eqref{funddec} in the presence of all the other eigenvalues of the ``background'' matrix $V$. This is the simplest possible holographic model, the holographic dimension being the ``eigenvalue'' space. In exactly the same way, in the decomposition \eqref{funddec}, the lower-right corners of the six elementary adjoint scalar fields in the $\nn=4$ theory in four dimensions yield the six emerging dimensions of the $\text{AdS}_{5}\times\text{S}^{5}$ dual geometry.

\smallskip

\noindent\emph{Remark}

\noindent There is actually a very important subtlety with the ``eigenvalue'' interpretation of the lower-right corner of the matrix $M$ in \eqref{funddec}, which is usually totally overlooked. This interpretation is only correct in a gauge for which the off-diagonal components of $M$ vanish. Such a gauge choice ``\`a la Landau'' is of course convenient for the one matrix model, but certainly not for full-fledged four dimensional gauge theories. When several matrices are present, it is actually impossible to find a gauge with an eigenvalue interpretation for all the matrices. Even for the one matrix model, it is useful and very instructive to compute $A(v)$ in other gauges \cite{onemmbranes}. 

This little discussion hints at the crucial importance of the partial gauge-fixing procedure in the construction of the holographic space. At the same time, it clarifies the situation in multi-matrix models. What could be the emerging space was always a bit puzzling in this context. Since the matrices a priori do not commute, the ``space of eigenvalues'' idea does not look promising. We now see that the same problem actually already occurs in the one matrix model in general gauges. The resolution of the puzzle is that the lower-right corners of all the matrices in the decomposition \eqref{funddec} can play the r\^ole of the emerging space, but the construction of this space is, in all cases, depending on a choice of partial gauge-fixing.

We shall say much more about the gauge-fixing in Sections \ref{gfsec} and \ref{dissec}. A detailed discussion for the one matrix model will be presented in \cite{onemmbranes}.

\subsubsection*{The pure Yang-Mills theory}

We can play the same game for the pure Yang-Mills D-brane system described by the actions \eqref{pureYMact}, \eqref{SmYM}. The set of adjoint Hermitian fields $\Phi$ of the form \eqref{phigenform} that we can build from the vector-like fields $W_{\mu}$ and $\bar W_{\mu}$ consists of an antisymmetric tensor, a traceless symmetric tensor and a scalar,
\be\label{phiYM} \bigl\{\Phi^{i}_{\ j}\bigr\} = \bigl\{ D_{\mu\nu},S_{\mu\nu},\phi \bigr\}\, .\ee
Explicitly, we define
\be\label{DSdef} D_{\mu\nu}  = \frac{1}{2i}\Bigl( 
W_{\mu}\bar W_{\nu} - W_{\nu}\bar W_{\mu}\Bigr)\, ,\quad 
S_{\mu\nu}  = \frac{1}{2}\Bigl( 
W_{\mu}\bar W_{\nu} + W_{\nu}\bar W_{\mu} 
- \frac{2}{d}\eta_{\mu\nu}W^{\rho}\bar W_{\rho}\Bigr)\ee
and
\be\label{phiYMdef}
\phi  = W^{\mu}\bar W_{\mu}\, .\ee
Note that if a parity violating $\theta$ angle is included in the theory, it is necessary to separate, in Euclidean language, the self-dual and anti self-dual pieces of $D_{\mu\nu}$. This can be done straightforwardly but we shall not discuss this case here for simplicity.

From our general discussion, we know a priori that the probe D-brane effective action will be a natural function of the $\uK$ adjoint fields $D_{\mu\nu}$, $S_{\mu\nu}$ and $\phi$ on top of the 
gauge field $A_{\mu}$. \emph{The scalar $\phi$ is naturally interpreted as describing one emerging holographic dimension, the long-sought fifth dimension of the dual string  description of the pure Yang-Mills theory.} The other fields $D_{\mu\nu}$ and $S_{\mu\nu}$ play r\^oles analogous to the variables $\La_{\alpha A}$ and $D_{\mu\nu}^{+}$ found in the more familiar context of the $\nn=4$ D-instanton, see \cite{fer1} and \eqref{phiinst}. They may be integrated out from the effective action by solving their classical equations of motion. Keeping them, however, may help to elucidate the strongly coupled physics in a more transparent way. For example, it has been suggested that models including an antisymmetric field $B_{\mu\nu}$ may be well-suited to describe confinement, see e.g.\ \cite{ellwang} and references therein. This offers clue that we may be on the right track here, providing a precise framework making links between holography and some of the ideas developed independently to understand the strongly coupled regime of gauge fields. We let the investigation of these fascinating issues for the future.

The form of the action $\hat S_{N,K}$, quadratic in the $W_{\mu}$ and $\bar W_{\mu}$ and linear in the Lagrange multipliers $L_{\mu\nu}$, $\ell_{\mu\nu}$ and $L$ implementing the constraints \eqref{DSdef} and \eqref{phiYMdef} can be derived straightforwardly, by using the identity
\be\label{genbilYM} W_{\mu}\bar W_{\nu} = S_{\mu\nu} + iD_{\mu\nu} + \frac{1}{d}\eta_{\mu\nu}\phi\ee
to express the quartic terms in \eqref{SmYM} in terms of $S_{\mu\nu}$, $D_{\mu\nu}$ and $\phi$. We find
\begin{multline}\label{hatSmYM} \hat S_{N,K}  =\frac{N+K}{\la} \int\!\d^{d}x\, \Biggl\{
-\nabla^{\mu}\bar W^{\nu}\bigl(\nabla_{\mu}W_{\nu}-\nabla_{\nu}W_{\mu}\bigr) +i W^{\mu}G_{\mu\nu}\bar W^{\nu}+ i \bar W^{\mu}F_{\mu\nu}W^{\nu}\\
+\frac{1}{4}\tr_{K}\biggl[ -3D^{\mu\nu}D_{\mu\nu} + S^{\mu\nu}S_{\mu\nu} - \frac{d-1}{d}\phi^{2} - 4 L\Bigl(\phi - W^{\mu}\bar W_{\mu}\Bigr)
\\- 4 L^{\mu\nu}\Bigl( D_{\mu\nu} - \frac{1}{2i}\bigl(W_{\mu}\bar W_{\nu}
- W_{\nu}\bar W_{\mu}\bigr)\Bigr) \\-4 \ell^{\mu\nu}\Bigl(S_{\mu\nu} - \frac{1}{2}\bigl( W_{\mu}\bar W_{\nu}
+ W_{\nu}\bar W_{\mu} -\frac{2}{d}\eta_{\mu\nu}\phi\bigr)\Bigr)\biggr]
\Biggr\} +\frac{K}{N}S_{N}(\mathscr V)\, .
\end{multline}
Using, in particular, the fact that $L_{\mu\nu}$ is antisymmetric and $\ell_{\mu\nu}$ symmetric and traceless, \eqref{hatSmYM} can be cast in the equivalent form
\begin{multline}\label{hatSmYM2} \hat S_{N,K}  =\frac{K}{N}S_{N}(\mathscr V)+\frac{N+K}{\la} \int\!\d^{d}x\, \Biggl[
-\nabla^{\mu}\bar W^{\nu}\bigl(\nabla_{\mu}W_{\nu}-\nabla_{\nu}W_{\mu}\bigr) \\+i W^{\mu}G_{\mu\nu}\bar W^{\nu} + \bar W^{\mu}(L\eta_{\mu\nu}+iL_{\mu\nu}+\ell_{\mu\nu})W^{\nu}
\\-\frac{1}{4}\tr_{K}\biggl[3D^{\mu\nu}D_{\mu\nu}-S^{\mu\nu}S_{\mu\nu}
+\frac{d-1}{d}\phi^{2}+ 4(L^{\mu\nu}-F^{\mu\nu})D_{\mu\nu}+4L\phi
+4\ell^{\mu\nu}S_{\mu\nu}\biggr]\Biggr]\, .
\end{multline}
In dimension $d\geq 4$, the technique of the Lagrange multipliers is made more complicated by renormalization. For example, in $d=4$, counterterms quadratic in $L_{\mu\nu}$, $\ell_{\mu\nu}$ and $L$ must be included in the action and thus the original interpretation of these fields as strict Lagrange multipliers will be lost. One may then try to use the analogue of the action \eqref{Sm1mmtilde} for the one-matrix model. This yields
\begin{multline}\label{tildeSmYM}
\hat S_{N,K} = \frac{N+K}{\la}\int\!\d^{d}x\, \Biggl[
-\nabla^{\mu}\bar W^{\nu}\bigl(\nabla_{\mu}W_{\nu}-\nabla_{\nu}W_{\mu}\bigr) \\
+i W^{\mu}G_{\mu\nu}\bar W^{\nu} 
+ \bar W^{\mu}\Bigl(-\frac{d-1}{2d}\eta_{\mu\nu}\phi+iF_{\mu\nu}-\frac{3}{2}iD_{\mu\nu}+\frac{1}{2}S_{\mu\nu}\Bigr)W^{\nu}
\\-\frac{1}{4}\tr_{K}
\biggl[S^{\mu\nu}S_{\mu\nu}-3D^{\mu\nu}D_{\mu\nu}
-\frac{d-1}{d}\phi^{2}\biggr]\Biggr]+\frac{K}{N}S_{N}(\mathscr V)\, .
\end{multline}
One can check straightforwardly that integrating out $S^{\mu\nu}$, $D^{\mu\nu}$ and $\phi$ from \eqref{tildeSmYM} yields \eqref{SmYM}, with the identifications \eqref{DSdef} and \eqref{phiYMdef}. One can also get \eqref{tildeSmYM} by integrating out $S_{\mu\nu}$, $D_{\mu\nu}$ and $\phi$ from \eqref{hatSmYM2} and renaming the remaining fields.

In the formulation using \eqref{tildeSmYM}, we run into the problem of the classical instability of the would-be emerging dimension $\phi$. However, unlike the case of the one matrix model, it is plausible that quantum corrections can stabilize the potential. For example, in $d=4$, it is natural to guess that a Coleman-Weinberg like term $\sim +\phi^{2}\ln\frac{\phi}{\La^{2}}$ will be generated. This would stabilize the probe branes at a dynamically generated scale $\phi\sim\La^{2}$, consistently with the creation of a mass gap in the model.

The action \eqref{tildeSmYM}, supplemented by appropriate ghost terms discussed in Section \ref{gfsec}, can be taken as the starting point to derive the pure Yang-Mills D-brane probe action, performing first the Gaussian integrals over $W_{\mu}$ and $\bar W_{\mu}$. Further discussion of this most interesting example will appear in separate publications.

\subsubsection*{Instantons in the $\nn=4$ theory\protect\footnote{This paragraph lies somewhat out of the paper's main line of development and may be omitted in a first reading.}}

As a last example, let us discuss D-instanton probes in the $\nn=4$ gauge theory. This is a case for which the probe branes and the background branes are different, along the lines explained in Section \ref{GenBsysSec}. We shall be brief, since a full treatment appeared previously in \cite{fer1}.

The action \eqref{probemixedinst} can be rewritten in such a way that the ADHM moduli with mixed indices $q,\bar q,\chi,\bar\chi$ appear only quadratically (see Eq.\ (3.11) and (3.14) of Ref.\ \cite{fer1}). To achieve this goal, one introduces six space-time scalars $\phi_{m}$ forming an $\text{SO}(6)$ vector, together with the Lagrange multipliers $D_{\mu\nu}^{+}$ and $\Lambda_{\alpha A}$ which implements the ADHM constraint ($D_{\mu\nu}^{+}$ is self-dual). All these new variables transform in the adjoint of $\uK$ and fit in a supersymmetry multiplet. They form the set of fields denoted by $\Phi$ in the previous paragraph,
\be\label{phiinst} \bigl\{\Phi^{i}_{\ j}\bigr\} = \bigl\{ \phi_{m},\Lambda_{\alpha A},D_{\mu\nu}^{+}\bigr\}\, .\ee
The effective probe action is a natural function of both $\Phi$ and the variables $\tilde M$ defined in \eqref{N4ivaviables}.

As shown in \cite{fer1}, the prescription \eqref{seffbetter3} yields the correct non-abelian action for D-instantons in the ten dimensional $\text{AdS}_{5}\times\text{S}^{5}$ geometry dual to the $\nn=4$ background brane theory. The variables $\phi_{m}$ correspond to the six emerging space dimensions, as in \cite{brit}. The full type IIB supegravity solution, including the metric and the self-dual Ramond-Ramond five-form field strength with the correct normalizations, can then be derived by comparing $A_{K}$ with Myers' non-abelian D-instanton action \cite{fer1}. This provides a non-trivial consistency check of our general ideas, albeit in a highly supersymmetric context. Generalizations along these lines were also worked out in \cite{fer3,fer4,fer5}.

The precise relation between the field $\phi_{m}$ and the mixed bilinears is given by
\be\label{phimvsbil} \phi_{m} = O^{-1}\cdot (v_{m}+\cdots)\, ,\ee
where the $\cdots$ denote terms which are non-zero when $\nn=4$ background fields $M$, see \eqref{N4ivaviables}, are turned on (these terms can be read off straightforwardly from eq.\ (3.14) in ref.\ \cite{fer1}). The operator $O$ and vector $v_{m}$ were defined in \eqref{Olindef} and \eqref{vmform} respectively. The relation \eqref{phimvsbil} is of the general form \eqref{phigenform}, albeit the dependence in the bilinears look rather complicated. Of course, the formula \eqref{phimvsbil} is suggested by the form of the action \eqref{probemixedinst} and is also highly constrained by supersymmetry: the fields $\phi_{m}$ so defined form a supersymmetric vector multiplet with the Lagrange multipliers $D_{\mu\nu}^{+}$ and $\Lambda_{\alpha A}$. 

Let us stress, however, that one may solve the model equally well by using the simpler bilinears of the form $q\bar q$ and $\chi\bar\chi$. For example, a particularly natural variable is the adjoint scalar
\be\label{sizeinst} s = q_{\alpha}\bar q^{\alpha}\, .\ee
The eigenvalues of $s$ are the moduli associated with the size of the instantons in the ADHM construction. If one includes both $s$ and the $\phi_{m}$ in the probe brane effective action $A_{K}$ (we shall explain below in Sec.\ \ref{sfsec} how to do so for any bilinear) and then integrates out $s$, one easily finds that
\be\label{sizevsrad} s\sim (\phi_{m}\phi_{m})^{-1}\, .\ee
This is the expected relation between the radial direction in $\text{AdS}_{5}$ and the size of instantons.

\section{Refinements and discussion}\label{refinesec}
\subsection{On the state/geometry mapping}\label{refinementsec}

The gauge theory action we start with may include sources for any gauge-invariant operator. This is equivalent to considering the theory in an arbitrary state. Turning on the sources does not change the discussion and the D-brane probe effective action can be constructed exactly as described above. The result of course will depend on the sources, or equivalently on the state of the gauge theory. The holographic geometry that we find is thus state-dependent, as expected. For example, one may consider the gauge theory at finite temperature, yielding in principle an effective action describing the motion of the probe branes in a black hole geometry.

\subsection{The partition function from the probe action\label{ZfromASec}}

Let us integrate Eq.\ \eqref{seffbetter} over $v$ and $\Phi$. Using \eqref{pathin1} and \eqref{stot0}, we find that
\be\label{ZNZNK} \int\!\mathcal D v\mathcal D \Phi \, e^{-\mathscr A_{N,K}(v,\Phi)} = \frac{Z_{N+K}}{Z_{N}}\, \cdotp\ee
This relation has a simple interpretation: the free energy of the $\text{U}(N+K)$ gauge theory can be obtained by adding the contribution from the $K$ D-brane probes to the free energy of the $\uN$ gauge theory.

Let us now consider \eqref{ZNZNK} in the leading large $N$ approximation. Since the action $\mathscr A_{N,K}\simeq NA_{K}$ is proportional to $N$, whereas the number of integration variables $v$ and $\Phi$ is of order $K^{2}$ and independent of $N$, the left-hand side can be evaluated by solving the classical equations of motion
\be\label{classicalAKmotion}\frac{\partial A_{K}}{\partial v} = \frac{\partial A_{K}}{\partial\Phi} = 0\ee
and plugging the solution into $A_{K}$. This yields the on-shell D-brane action $A_{K}^{*}$ and
\be\label{ZNZleft} \ln \int\!\mathcal D v\mathcal D \Phi \, e^{-\mathscr A_{N,K}(v,\Phi)} = -NA_{K}^{*} + O(1)\, .\ee
As for the right-hand side of \eqref{ZNZNK}, we can use the known form of the large $N$ expansion of the partition function given by \eqref{ZNexp} to derive that
\be\label{ZNZright} \ln\frac{Z_{N+K}}{Z_{N}} = -(N+K)^{2}F^{(0)} + N^{2}F^{(0)}+ O(1) = -2NKF^{(0)} + O(1)\, .\ee
The crucial point of this simple equation is that the terms of order $N$ involve only the planar partition function $F^{(0)}$. Comparing \eqref{ZNZleft} with \eqref{ZNZright}, we obtain the fundamental relation
\be\label{ZArel} F^{(0)} = \frac{1}{2K}A_{K}^{*}\ee
which was already mentioned in Section \ref{genDpSec}. Since $F^{(0)}$ does not depend on $K$, we also find \eqref{AKAonshell} and \eqref{fundrel2}.

Eq.\ \eqref{ZArel} or \eqref{fundrel2} are crucial for the whole formalism, because they show that the probe brane effective action $A_{K}$ contains all the information about the large $N$ gauge theory, even at $K=1$, and can thus be used as an alternative to more standard functionals, like the 1PI effective action $\Gamma$, to study the gauge theory. \emph{Possibly the main lesson of the present paper is to suggest that focusing on $A_{K}$ is much more fruitful and promising to understand the strongly coupled physics, because the holographic description of the theory then becomes manifest from first principles.}

Of course, by starting from \eqref{ZNZNK}, which is valid for any finite $N$ and $K$, one can generalize the relation \eqref{ZArel} to subleading orders in $1/N$. On the one hand, the full $1/N$ expansion of the right-hand side of \eqref{ZNZNK} is obtained straightforwardly from \eqref{ZNexp}. On the other hand, the $1/N$ corrections to the left-hand side are obtained by taking into account the $1/N$ corrections in the action $\mathscr A_{N,K}$, which are given by the expansion \eqref{AKNexp}, together with quantum loop corrections computed from this action. The loop counting parameter is $1/N$, since $\mathscr A_{K,N}$ grows like $N$ at large $N$.

\subsection{The open string field theory}\label{sfsec}

One of the main point made in Section \ref{diagsec} is that the probe brane effective action depends not only on the variables $v$ but also crucially on the adjoint bilinears $\Phi$ of the form \eqref{phigenform}. But this is not the end of the story. One can consider any variable of the form
\be\label{SFvar} \Phi(\mathscr O)^{i}_{\ j} = w^{i}_{\ a}\mathscr O^{a}_{\ b}\bar w^{a}_{\ j}\, ,\ee
where $\mathscr O^{a}_{\ b}$ is any operator in the background brane theory transforming in the adjoint of $\uN$. The fields $\Phi(\mathscr O)$ are all in the adjoint of $\uK$. The variables that we have emphasized in Section \ref{diagsec} correspond to the special case where the operator $\mathscr O$ is the identity. For example, in the one matrix model at large $N$, a full set of independent variables is obtained by considering
\be\label{phinmm} \Phi(V^{n})=\phi_{n} = wV^{n}\bar w\, .\ee
These operators are all of order $O(1)$ at large $N$ and the variable $\phi$ defined in \eqref{phidef1mm} corresponds to $\phi_{0}$.

From the point of view of the probe branes, the operators $\Phi(\mathscr O)$ form an infinite set of independent fields. They are naturally associated with the infinite tower of open string states of the open string theory describing the dynamics of the probe D-branes in the holographic geometry generated by the background branes. So we get a map from any adjoint operator on the background branes to open string states on the probe branes. This map is actually one-to-many, since there are is general several different mixed fields $w,\bar w$ that can be used in \eqref{phinmm}. This is the analogue of the mapping between the gauge invariant operators of the background brane theory and the closed string states of the dual closed string theory.

It is very easy to integrate in all the fields $\Phi(\mathscr O)$ to get the string field theory action describing the probe D-branes. One way to do this is to introduce sources $J_{\mathscr O}$ for all the operators $\Phi(\mathscr O)$ in the definition \eqref{seffbetter} and then Legendre transform the resulting functional with respect to the sources. Another procedure is to use Lagrange multipliers, as we have done in Section \ref{exsec4sec}, to integrate in the variables $\Phi$. At large $N$, these two procedures turn out to be strictly equivalent.

Let us give the details in the case of the one matrix model. The ``string field'' is denoted by $\Phi = \{\phi_{n}=wV^{n}\bar w\, ,\ n\geq 0\}$. The method of Lagrange multipliers amounts to writing
\begin{multline}\label{SeffSF} e^{-\mathscr A_{N,K} (v,\Phi)} = \frac{e^{-(\frac{N}{K}+1)S_{K}(v)}}{Z_{N}}\int\!\mathcal D V \mathcal D w
\mathcal D \bar w \prod_{n\geq 0}\mathcal D L_{n}\\
e^{-S_{N}(V) -S_{N,K}(V,v,w,\bar w) + \frac{N+K}{\la}\sum_{n\geq 0}\tr_{K}
L_{n}(w V^{n}\bar w - \phi_{n})}\, ,
\end{multline}
where the Lagrange mulpliers $L_{n}$ are in the adjoint of 
$\uK$.\footnote{As explained in Section \ref{diagsec}, $S_{N,K}$ can be replaced in \eqref{SeffSF} by an action $\hat S_{N,K}$ which is quadradic in $w$ and $\bar w$ but depends on $\phi_{0}=\phi$, see Eq.\ \eqref{Smhat1mm}.} Introducing the standard generating function $\mathsf W$ of the connected correlation functions of the operators \eqref{phinmm}, we can rewrite \eqref{SeffSF}, in the leading large $N$ approximation, as
\be\label{SeffSF2} e^{-N A_{K} (v,\Phi)} =\int\! \prod_{n\geq 0}\mathcal D L_{n}\, e^{\mathsf W(L_{0},L_{1},\ldots) -\frac{N}{\la}\sum_{n\geq 0}\tr_{K}L_{n}\phi_{n}}\, .\ee
Following the discussion of the large $N$ counting in Sec.\ \ref{diagsec}, it is clear that $\mathsf W$ is proportional to $N$ in the 't~Hooft's scaling. At large $N$, the integral over the Lagrange multipliers can thus be done by solving the saddle point equations
\be\label{saddleST} \frac{\partial}{\partial L_{n}} \mathsf W\bigl(L_{0},L_{1},\ldots\bigr) = \frac{N}{\la}\phi_{n}\, .\ee
Plugging the solution back into \eqref{SeffSF2} then amounts to performing the Legendre transform of the generating function $\mathsf W$ with respect to the sources. This Legendre transform thus coincides with the string field theory action $NA_{K}(v,\Phi)$, as expected. Let us note that since $NA_{K}$ can be treated classically when $N\rightarrow\infty$, one can integrate out any number of fields $\phi_{n}$ simply by solving the corresponding classical equations of motion. The D-brane effective action discussed in Section \ref{diagsec} is recovered after all the fields except $\phi=\phi_{0}$ are integrated out in this way.

\subsection{On the number of space-time dimensions}\label{spacedimsec}

The discussion of the previous subsection forces us to re-examine the notion of holographic space dimensions. At the end of Section \ref{treesumsec}, we proposed to identify the emergent dimensions with the adjoint bilinears $\sim w\bar w$ that are \emph{scalars} with respect to the Lorentz group acting on the background branes. For example, for the pure Yang-Mills theory, we had one such scalar given by Eq.\ \eqref{phiYMdef}. However, it is actually natural to consider an infinite number of scalars, of the form \eqref{SFvar}, which can be built using the $\uN$ field strength. For example,
\be\label{manyscalarsYM} W^{\mu}G_{\mu\nu}\bar W^{\nu}\, ,\quad
W^{\mu}G^{\nu\rho}G_{\nu\rho}\bar W_{\mu}\, ,\quad W^{\mu} G_{\mu\rho}G^{\rho\nu}\bar W_{\nu}\, ,\quad \text{etc.,}\ee
all correspond to scalar open string states living on the probe branes. The simplest bilinears \eqref{phiYMdef}, which are associated with the identity operator in the background brane theory, are in a sense special because they help summing up the trees of bubble diagrams, as explained in Section \ref{diagsec}. However, there is no fundamental distinction with all the other scalar excitations like \eqref{manyscalarsYM}. One might thus consider that the probe branes see an infinite number of emerging dimensions, one for each scalar open string state living on the branes.

This ambiguity is related to the fact that the emerging dimensions are typically not large, a point already raised in Section\ \ref{IntroSec}. From the point of view of the probe branes, it is natural to keep only the low mass modes on the branes in an effective description. This is unambiguous when there is a parameter in the model that allows to separate the mass scales between (approximately) massless degrees of freedom, the scalars amongst these playing the r\^ole of the space dimensions, and highly massive excitations that can be integrated out. In models like the pure Yang-Mills theory, where there is only one physical scale $\La$, the distinction between the fifth dimension \eqref{phiYMdef} and other more massive scalar excitations cannot be sharp. We may still expect that a five dimensional model would offer a correct and simple description of the physics in some regime, but in general other modes, including scalars, will have to be included. Similarly, the use of a local expansion for the probe branes effective action will only be a rough approximation. An accurate description requires the use of the non-local form of the action, which, in the string language, amounts to including all the $\alpha'\sim 1/\La^{2}$ corrections.

As an illustration of the ambiguity in the number of space dimensions seen by the probe branes, let us consider the familiar case of the pure $\nn=4$ Yang-Mills model, which we deform by adding a mass $m$ to all the matter fields. This breaks supersymmetry down to $\nn=0$. When $m$ is very large, we expect a transition between the ten dimensional $\text{AdS}_{5}\times\text{S}^{5}$ geometry and a very different description adequate for the pure Yang-Mills model, involving a five-dimensional bulk together with excited string modes. How is this transition most likely seen from the point of view of probe D3-branes?

\noindent (i) When $m=0$, the only scales in the probe action are the string length $\ls$ and the radius $R\sim \ls\la^{1/4}$ of $\text{AdS}_{5}$, where $\la$ is the 't~Hooft coupling. If $\la\gg 1$, there is a sharp separation between, on the one hand, low energy modes, governed by the scale $1/R$ and described by the variables $v$ in the decomposition \eqref{funddec} and, on the other hand, excited open string states, governed by the scale $1/\ls$ and described by the variables $\Phi$ defined in \eqref{SFvar}. A good effective description of the probe branes is thus obtained by keeping only $v$. Since there are six scalars $\phi_{m}$, $1\leq m\leq 6$, in $v$, corresponding, in the decomposition \eqref{funddec}, to the lower-right corner of the six massless elementary scalars of the $\nn=4$, $\text{U}(N+K)$ gauge theory, there are six dimensions transverse to the probe branes. The probe brane effective action, computed along the lines explained in the previous sections and in which all the fields except the variables $v$ have been integrated out, should thus be reliably approximated by the standard Dirac-Born-Infeld plus Chern-Simons D-brane non-abelian action on the $\text{AdS}_{5}\times\text{S}^{5}$ background.

The open string scalar mode $\phi$ defined in \eqref{phiYMdef} is of course present, as in the pure Yang-Mills case. This mode will have an overlap with the excited states of mass $\sim 1/\ls$. It is natural to suspect that it will also have an overlap with the operator $\phi_{m}\phi_{m}$ (possibly vanishing when $\la\rightarrow\infty$), which is associated with the radial direction in $\text{AdS}$ and has the same quantum numbers. If we integrate it out, we expect to find a relation of the form $\phi\sim\phi_{m}\phi_{m}$. This mechanism would be similar to the relation \eqref{sizevsrad} in the D-instanton case.

\noindent (ii) When $m\not = 0$, we get a dynamically generated mass scale $\La$ on top of the other scales $m$, $R\sim \ls\la^{1/4}$ and $\ls$. The low energy physics is governed by the pure Yang-Mills theory when the $\nn=4$ matter fields decouple, which implies that $m\gg\La$. In this limit,
\be\label{dyngen} \La \sim m\,  e^{-\frac{48\pi^{2}}{11\la}}\, ,\ee
the numerical factor in the exponential being related to the pure Yang-Mills beta function. This relation shows that the decoupling requires $\la\rightarrow 0$. In this regime, there is no longer a separation of scales and the excited string modes play a r\^ole. The ``elementary'' scalars $\phi_{m}$ become extremely massive, even more massive than the typical excited open string states, whereas the mass of the ``composite'' scalar $\phi$ will remain of order $\La$. Whereas when $m\ll\La$, it was natural to integrate out $\phi$ and keep the $\phi_{m}$, in the opposite limit $m\gg\La$ the most natural description of the probe brane theory is obtained by keeping $\phi$ and integrating out the $\phi_{m}$. This yields the expected five dimensional bulk description. The foregoing caveat must of course always be kept in mind: a precise description will also involve other excited string modes, which can never be completely decoupled.

\section{The bare bubble approximation}\label{bareapsec}

The \emph{exact} calculation of the probe brane effective action is out of reach of present technology except in very special and simplified models, like the one matrix model in zero dimension \cite{onemmbranes}. Even in highly supersymmetric models, only a few terms in the action will be protected. Computing these terms using the many powerful mathematical tools at our disposal in the supersymmetric context (holomorphy, matrix models, localization, etc.) could certainly be a very fruitful direction of research to pursue. However, our main interest goes towards more realistic theories which are not amenable to study using these techniques.

One of the most engaging aspect of the approach we are developing is that it suggests in a very natural way a new non-perturbative approximation scheme which can be applied to any gauge theory, independently of supersymmetry, and including in particular the case of the pure Yang-Mills model. The idea of the approximation comes directly from the discussion in Section \ref{diagsec}. There we showed that the computation of the probe brane effective action can be decomposed into two steps. In the first step, one computes the dressed bubbles; in the second step, one puts the dressed bubbles together to form trees and sum up the trees. This last step can always be performed exactly, using the vector model techniques. The idea of the approximation is then to take into account the dressing of the bubbles only approximately, whereas the sum over trees is done exactly, using the approximate bubbles. The ``bare bubble'' approximation amounts to using the simplest possible approximation for the bubbles, in which only the first one-loop term in the expansion depicted in Fig.\ \ref{dressedbubble} is kept. This is equivalent to discarding the quantum corrections coming from the integration over the background brane fields in \eqref{seffbetter} or, equivalently, to setting the background brane fields $V$ to their classical values. For example, for the pure Yang-Mills theory in the vacuum state, we would simply set the $\uN$ gauge field to zero, $\mathscr V_{\mu} =0$; in a thermal state, we would use the blackbody radiation field; etc.

It is important to realize that this ``bare bubble'' approximation scheme is very different from the usual perturbative expansion. Indeed, the sum over the trees of bubbles always produce an infinite series over the 't~Hooft coupling $\la$, even if the simplest one-loop form for the bubbles is used. The leading bare bubble approximation is thus non-perturbative in nature. 

\begin{figure}
\centerline{\includegraphics[width=6in]{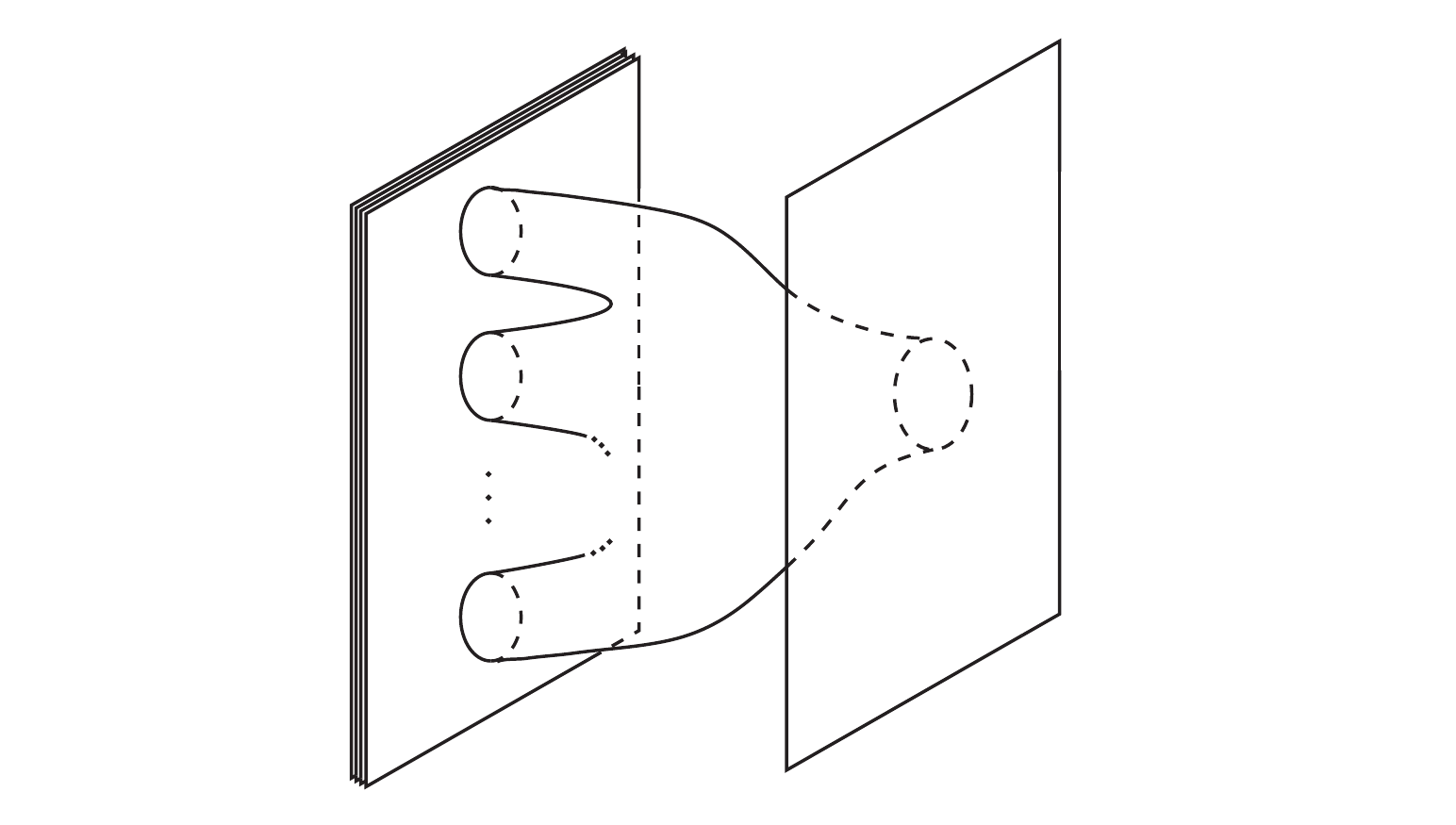}}
\caption{Worldsheet diagram corresponding to a leading large $N$ Feynman diagram, as the one depicted in Fig.\ \ref{typicalDfig}. The separation between the branes allows to distinguish clearly the background branes, on the left, and the probe branes, on the right, but is otherwise immaterial. The number of worldsheet boundaries on the background branes correspond to the number of loops in the Feynman diagrams. There is only one boundary on the probe branes and no holes in the worldsheet.\label{worldsheetfig}}
\end{figure}

This is made particularly clear by using the worldsheet representation of the Feynman diagrams. A typical worldsheet contributing to the large $N$ probe brane effective action is depicted in Fig.\ \ref{worldsheetfig}. The unique boundary on the probe branes corresponds to the unique dashed line closed loop in the typical Feynman diagram drawn in Fig.\ \ref{typicalDfig}. The boundaries on the background branes correspond to the plain line closed loops. In the bare bubble approximation, the number of boundaries is equal to the number of bubbles in the diagram, which can be arbitrary. We thus see that \emph{the bare bubble approximation includes diagrams with all the possible topologies,} having an arbitrary number of boundaries on the background branes. To the contrary of the standard perturbation expansion, which is a truncation in the topology of the diagrams, the bare bubble approximation rather truncates the moduli space of the possible string diagrams. Only the diagrams for which the number of boundaries on the background branes equal the number of bubbles are kept.

By construction, the bare bubble approximation is always one-loop exact. It will thus always capture correctly the weak coupling physics, for example asymptotic freedom. Moreover, the non-perturbative nature of the approximation implies that it may be able to capture, at least qualitatively, some important aspects of the non-perturbative strongly coupled physics at the same time. Having at hand a computational method able to cover all regimes of a gauge theory is an exciting novelty. Of course, detailed studies in specific models need to be performed before the reliability of the approximation can be assessed.

At the moment, we can only report on the one matrix model in zero dimension with a quartic potential, which is worked out in full details in \cite{onemmbranes}. The results are very encouraging. In the bare bubble approximation, the computation of the probe brane effective action is a back-of-an-envelope calculation in this case. By using \eqref{fundrel2}, we then derive an approximate free energy which turns out to match the known exact result with a remarkably good accuracy, both at weak and strong coupling. Actually, the error made never exceeds $3\%$ uniformly in the 't~Hooft coupling, all the way from the perturbative small $\la$ regime to the strong coupling $\la\rightarrow\infty$ regime!

The bare bubble approximation can be upgraded in two ways. An obvious and systematic method is to dress the bubbles perturbatively: instead of keeping only the one-loop diagram in the expansion of Fig.\ \ref{dressedbubble}, we could sum all the diagrams up to $L$ loops (and, of course, still sum exactly all the trees of the bubbles so obtained). This will clearly improve the results at small or even moderate values of the 't~Hooft coupling. Another, more promising way to improve significantly the results would be to compute the bubbles in some strong coupling approximation/numerical scheme, for example on the lattice. Since the emergence of the holographic dimensions is understood from the sum over the trees of bubbles, which is exact, this point of view allows in principle to merge the lattice techniques (used to evaluate the dressed bubbles) with the holographic description. It would be extremely interesting to investigate the picture that emerges along these lines for the pure Yang-Mills theory or QCD.

Let us finally suggest a simple physical interpretation for the bare bubble approximation. As we have already pointed out, from the microscopic point of view, using bare bubbles amounts to treating the background brane fields $V$ classically. So we can interpret the bare bubble approximation as a sort of mixed scheme, in which the background branes are treated classically whereas the microscopic interactions between the probe and the background branes are treated quantum mechanically and exactly. As we have seen, the emergence of a classical holographic description from the point of view of the probes can be understood from the latter quantum effects only.

Of course, the effective interactions between probes and background are modified by the quantum nature of the background branes but, intuitively, these corrections should be small ``far away'' from the background branes. ``Far away'' from the sources means, in the dual geometry, at large radial direction. For example, in the pure Yang-Mills theory, this would correspond to large $\phi$. This region probes the UV of the background brane theory. This is a manifestation of the usual UV/IR relation and can be seen, for example, from the action \eqref{tildeSmYM}; large $\phi$ corresponds to a large mass for the off-diagonal fields $W$ and $\bar W$, which play the r\^ole of $W$-bosons. Asymptotic freedom then implies that the bare bubble approximation should certainly be reliable in this case. The situation is somehow analogous to the classical treatment of the stress energy tensor in general relativity. Under a wide range of circumstances, the quantum nature of the source of the gravitational field is not expected to modify relevantly the space-time geometry. This is particularly true far away from the sources, but it could be quite reliable even ``in the IR,'' for example near the horizon of a large black hole. 

More precisely, the physics is extracted from the holographic description by going on-shell, as \eqref{fundrel2} shows. The position of the branes in the emergent geometry is thus determined by solving their classical equations of motion. The solution of course depends on which sources are turned on, i.e.\ on which question in the gauge theory we focus on. If we are probing the UV behaviour, then the branes will sit in the UV region of the geometry. If we are studying the vacuum state, they will probe deeper in the bulk and corrections to the bare bubble approximation may then be important. However, we do think very plausible that the bare bubble approximation could describe correctly some aspects of the physics in this regime as well, including in the pure Yang-Mills theory. After all, as we have emphasized, it does treat exactly some crucial non-perturbative aspects of the quantum physics. For example, in the one matrix model, the deep bulk geometry corresponds to the support of the density of eigenvalues of the background branes. It turns out that the on-shell probe branes do sit on or near this deep bulk region \cite{onemmbranes}. In spite of this fact, the bare bubble approximation is very successful.

\section{Gauge fixing} \label{gfsec}
\subsection{General discussion}\label{gdgfsec}

We are now going to fill a last important gap left open in the preceding sections, by carefully explaining the gauge-fixing procedure that must be used to define the probe D-brane effective action. We take up this point only now because it does not interfere in any essential way with our foregoing discussion. However, it is an absolutely crucial and surprisingly subtle aspect of the whole framework.

The basic point is as follows. A system of $N+K$ branes is described microscopically by a $\text{U}(N+K)$ gauge theory, whereas two stacks of $N$ and $K$ branes are more naturally described by a $\text{U}(N)\times\text{U}(K)$ gauge theory. Moreover, when $N\rightarrow\infty$, we wish to replace the background branes by the dual closed string background; the natural description is then in terms of a $\uK$ gauge theory. Let us emphasize that this ``symmetry breaking'' $\text{U}(N+K)\rightarrow\uN\times\uK\rightarrow\uK$ has nothing to do with a Brout-Englert-Higgs mechanism. It is very important to understand this point. For example, the models we consider may not even have scalar fields. The reduction of the gauge group is rather associated with \emph{equivalent reformulations} of the model. This is possible because the gauge symmetry is a redundancy in the description and not a real physical symmetry. Technically speaking, the definition of the background plus probe D-brane system requires to \emph{partially gauge fix the original gauge symmetry $\text{U}(N+K)$ down to $\uN\times\uK$} (the further gauge fixing down to $\uK$ is trivial since it amounts to a standard full gauge fixing of the $\uN$ factor). 

At this stage, there is a risk of a potentially very prejudicial confusion, which seems to be widespread in the previous literature on D-branes, between two completely different notions: partial gauge-fixing on the one hand and the background field gauge on the other hand. Since the distinction is crucial to understand the correct construction of the D-brane actions, let us try to clarify the difference straight away. So let us consider a gauge theory with gauge group $G$ and pick a subgroup $H\subset G$. In our case, $G=\text{U}(N+K)$ and $H=\uK$ (or $\uN\times\uK$).

In a background field gauge formalism, \emph{the gauge symmetry $G$ is completely fixed}. However, the gauge fixing conditions for $G$ can be parameterized by a classical background gauge field in such a way that a new gauge symmetry $H_{\text{back.}}$ appears in the problem. Group theoretically, $H_{\text{back.}}$ can be isomorphic to any subgroup of $G$, by chosing appropriately the background gauge field and the gauge fixing conditions. In most applications, $H_{\text{back.}}$ is actually isomorphic to $G$ itself. Having this new gauge invariance at hand can be very useful to simplify and organize certain calculations, as is well-known. In the D-brane context, one may be tempted to use this formalism to build a 1PI effective action invariant under a background field gauge invariance $H_{\text{back.}}$ isomorphic to $\uK$ (or $\uN\times\uK$) and identify this effective action with the D-brane action. \emph{We claim that this is not the correct prescription to define a D-brane action}.

A simple way to understand the problem is to realize that the background field gauge invariance is not part of the original gauge symmetry, but an auxiliary, purely classical gauge symmetry that is introduced for technical convenience. This is unlike the $\uK$ gauge symmetry associated with the $K$ probe branes (or the $\uN\times\uK$ gauge symmetry associated with two stacks of $N$ and $K$ branes), which must be part of the original quantum gauge symmetry, because the $K$ branes are part of the full stack of $N+K$ branes on which the $\text{U}(N+K)$ gauge symmetry lives. Another way to reach the same conclusion is to note that the 1PI effective action defined from the background field gauge is a purely classical object, encoding the full quantum corrections to the original gauge theory, even for finite $N$. On the other hand, the D-brane action is expected to describe fluctuating quantum degrees of freedom living on the branes, as for example the fluctuating target space coordinates. The conclusion is that our D-brane action $\mathscr A_{N,K}$ \emph{is not a 1PI effective action computed in a background field gauge}.

The correct procedure to define the D-brane probe action is to use instead a \emph{partial gauge-fixing} of the original gauge symmetry, down to the subgroup $H=\uK$. Unlike the gauge symmetry found in the background field gauge, the subgroup that remains unbroken after the partial gauge-fixing is part of the original gauge symmetry acting on the quantum fields. At finite $N$, the resulting $H=\uK$-invariant action must be treated quantum mechanically, with a loop counting parameter proportional to $1/N$, as it should. 

The partial gauge-fixing procedure turns out to be quite interesting in itself and has been developed in full generality and from first principles in \cite{ferequiv}. This construction will be briefly reviewed below. Relevant references can also be found in the lattice literature \cite{Schaden,Shamir}, the prime motivation in this context being to find a non-perturbative formulation of the gauge theory, but still trying the fix the gauge as much as possible. 

\subsection{Partial gauge fixing and equivariant cohomology}\label{equivsec}

Consider a gauge theory with gauge group $G$, $G$-invariant action $S$, fields $M$ and partition function $Z$. The action $S$ may include sources for all gauge-invariant operators, in which case $Z$ encodes all the physical information about the theory. The partition function is given by a path integral of the form
\be\label{Zdef} Z = \int\! \mathcal D M \mathcal D[\text{ghosts}]_{G}\, e^{-S(M) + s_{G}\psi_{G}}\, ,\ee
where $[\text{ghosts}]_{G}$ is a set of ghost fields suitable to gauge-fix $G$, $\psi_{G}$ is a gauge-fixing fermion depending on the ghosts and $s_{G}$ the standard nilpotent BRST operator for the gauge group $G$. Of course, the partition function does not depend on the gauge choice, i.e.\ it does not depend on $\psi_{G}$.

We would like to reformulate the model in terms of a gauge group $H\subset G$. We split the set of fields $M$ into two subsets,
\be\label{Vsplit}M = \{\mathcal V,\mathcal W\}\, ,\ee
in such a way that $\mathcal V$ and $\mathcal W$ each belong to a representation of $H$ and thus do not mix under a gauge transformation belonging to $H$. The splitting \eqref{Vsplit} is of course not unique and the construction can be done for any choice of splitting. For example, for our D-brane system with $G=\text{U}(N+K)$ and $H=\uN\times\uK$, we choose
\be\label{specialsplit} \mathcal V = \bigl\{V,v\bigr\}\, ,\quad\mathcal W = \bigl\{w,\bar w\bigr\}\, ,\ee
where the fields $V,v,w,\bar w$ are defined by the matrix decomposition \eqref{funddec}.

The goal is then to define a $H$-invariant action $S_{H}$ depending on $\mathcal V$ such that the partition function of the gauge theory of gauge group $G$ and action $S$ can be equivalently written as the partition for the gauge theory of gauge group $H$ and action $S_{H}$,
\be\label{ZHdef} Z = \int\! \mathcal D\mathcal V \mathcal D[\text{ghosts}]_{H}\, e^{-S_{H}(\mathcal V) + s_{H}\psi_{H}}\, ,\ee
the $[\text{ghosts}]_{H}$, $s_{H}$ and $\psi_{H}$ representing standard ghost fields, BRST operator and gauge-fixing fermion suitable to gauge fix $H$. 

Intuitively, $S_{H}$ is defined by integrating out the fields $\mathcal W$ and the ghosts associated to the ``broken generators'' belonging to $G/H$ in the formula \eqref{Zdef}. However, the precise implementation of this idea is not straightforward and involves some interesting subtleties explained in details in \cite{ferequiv}. The result is that the possible actions $S_{H}$ are given by a path integral of the form 
\be\label{SHdef} e^{-S_{H}[\psi_{G/H}](\mathcal V)} = \int\!\mathcal D \mathcal W\mathcal D[\text{ghosts}]_{G/H}\, e^{-S(\mathcal V,\mathcal W) + \delta\psi_{G/H}}\, .\ee
In this formula, $[\text{ghosts}]_{G/H}$ is a set of ghost fields belonging to the quotient $\mathfrak g/\mathfrak h$ of the Lie algebras of $G$ and $H$, $\psi_{G/H}$ is a partial gauge-fixing fermion which is invariant under gauge transformations belonging to $H$ and $\delta$ is an equivariant differential of ghost number one associated with the equivariant cohomology of $G$ with respect to $H$. In particular, $\delta$ is not nilpotent but squares to a $H$ gauge transformation. Detailed formulas for the action of $\delta$ and possible fermions $\psi_{G/H}$ are given in \cite{ferequiv}. Eq.\ \eqref{SHdef} has the following fundamental properties \cite{ferequiv}:

\noindent (i) It defines an action $S_{H}[\psi_{G/H}]$ which is manifestly invariant under gauge transformations belonging to $H$.

\noindent (ii) As the notation indicates, the action $S_{H}[\psi_{G/H}]$ does depend on the choice of the gauge-fixing fermion $\psi_{G/H}$. More precisely, it depends on $\psi_{G/H}$ modulo the addition of $H$-invariant terms of ghost number minus one that are $\delta$-closed. However, a central consequence of the analysis in \cite{ferequiv} is to show that the partition function defined by \eqref{ZHdef} does not depend on $\psi_{G/H}$, nor of course on $\psi_{H}$ as a consequence of the $H$-invariance of $S_{H}$. Moreover, it coincides with the partition function defined by \eqref{Zdef}. In this sense, all the actions $S_{H}[\psi_{G/H}]$ are physically equivalent.

\noindent (iii) The gauge-fixing Lagrangian obtained by evaluating $\delta\psi_{G/H}$ always contains quartic ghost term, \emph{even at tree level.} This is a fundamental difference with the standard gauge-fixing procedure of the full gauge group which can be done with a quadratic ghost Lagrangian. The only exceptions to this rule is when $G/H$ is a group, in which case the partial gauge fixing is trivial and reduces to the standard gauge fixing of the group $G/H$, or when one uses a gauge choice \`a la Landau imposed via a delta-function in the path integral. 

Let us comment a bit further on this last point. It is well-known that in gauge theories, quartic ghost counterterms must be included if a non-linear gauge choice is made, in order to define the renormalized 1PI effective action. These quartic ghost terms are BRST-exact and thus do not contribute to gauge invariant observables. They are needed only when focusing on the 1PI effective action, which is not gauge invariant, but yet encodes in a convenient way the gauge-invariant information about the theory. The quartic ghost terms appearing in the partial gauge-fixing procedure are of an entirely different nature. They appear at \emph{tree-level} and are unrelated a priori to renormalization. For example, they must be included even in zero dimensional models like the matrix model studied in \cite{onemmbranes}. To the opposite of the familiar quartic ghost terms which do not contribute to gauge invariant quantities, they are actually essential to ensure the gauge invariance of the procedure. Without them, the partition function defined by \eqref{ZHdef} would actually do depend on $\psi_{G/H}$! The matrix model provides a beautiful illustration of these properties \cite{onemmbranes}.

There is an interesting twist to this story that is useful to mention. Quartic ghost terms have been introduced before in studies of the Abelian projection \cite{Kondo}, in order to improve the renormalization properties of the 1PI action defined in this way. This was seen as an improvement over more standard Maximal Abelian Gauge choices. However, the Abelian projection \cite{tHooftAP} is actually an example of a partial gauge-fixing, from a gauge group $\uN$ down to $\text{U}(1)^{N}$. Its correct treatment should thus always be done in the formalism of \cite{ferequiv} and thus \emph{must} include tree-level quartic ghost terms! We shall discuss further some analogies between the D-brane picture developed in the present paper and some aspects of the Abelian projection scenario in Section \ref{APSec}.

\subsection{The complete definition of the probe D-brane action}

In the light of the previous subsection, let us now sum up all the steps in the construction of the probe D-brane action $\mathscr A_{N,K}$, modifying the discussion of Section \ref{treesumsec} to take into account the gauge-fixing.

\noindent\emph{Step one}: Define the action $S_{N,K}[\psi]$ by adding to \eqref{stot0} an equivariant gauge-fixing term $\delta\psi$ taking care of the partial gauge fixing of $\text{U}(N+K)$ down to $\uN\times\uK$,
\be\label{stotfin} S_{N+K}(M)-\delta\psi = S_{N}(V) + \Bigl(\frac{N}{K}+1\Bigr)S_{K}(v) + S_{N,K}[\psi](V,v,w,\bar w,\eta,\bar\eta,\chi,\bar\chi)\, .\ee
The variables $\eta^{i}_{\ a}$, $\bar\eta^{a}_{\ i}$ and $\chi^{i}_{\ a}$, $\bar\chi^{a}_{\ i}$ are the ghosts and anti-ghosts respectively, belonging to $\mathfrak u(N+K)/\mathfrak u(N)\oplus\mathfrak u(K)$. 

\noindent\emph{Step two}: Introduce new variables $\Phi$ such that
\be\label{pathI1} e^{-S_{N,K}[\psi](V,v,w,\bar w,\eta,\bar\eta,\chi,\bar\chi)} = \int\!\mathcal D\Phi\, e^{-\hat S_{N,K}[\psi](V,v,\Phi,w,\bar w,\eta,\bar\eta,\chi,\bar\chi)}\, ,\ee
where the action $\hat S_{N,K}[\psi]$ is \emph{quadratic} in the mixed fields. This is always possible, by using variables 
\be\label{phigenf2} \bigl\{\Phi^{i}_{\ j}\bigr\} \sim \bigl\{w^{i}_{\ a}\bar w^{a}_{\ j}\, , 
w^{i}_{\ a}\bar\eta^{a}_{\ j}\, , \eta^{i}_{\ a}\bar w^{a}_{\ j}\, , w^{i}_{\ a}\bar\chi^{a}_{\ j}\, , \chi^{i}_{\ a}\bar w^{a}_{\ j}\, ,
\eta^{i}_{\ a}\bar\eta^{a}_{\ j}\, ,\chi^{i}_{\ a}\bar\chi^{a}_{\ j}\, ,\eta^{i}_{\ a}\bar\chi^{a}_{\ j}\, , \chi^{i}_{\ a}\bar\eta^{a}_{\ j}
\bigr\}\ee
in the adjoint of $\uK$ which are bilinears in the mixed fields $w,\bar w,\eta,\bar\eta,\chi,\bar\chi$. Note that, on top of the bilinears of the form \eqref{phigenform}, bilinears involving the ghosts need to be introduced, in order to make quadratic the quartic ghost couplings which are always present in the partial gauge-fixing (except if a Laudau-like gauge is used).

\noindent\emph{Step three}: Define the probe D-brane action $\mathscr A_{N,K}[\psi]$ by
\begin{multline}\label{ANKdef1} e^{-\mathscr A_{N,K}[\psi](v,\Phi)} = \frac{e^{-(\frac{N}{K}+1)S_{K}(v)}}{Z_{N}}\int\!\mathcal D V\mathcal D[\text{ghosts}]_{\uN}\mathcal D w\mathcal D\bar w \mathcal D\eta\mathcal D\bar\eta\mathcal D\chi\mathcal D\bar\chi\\ e^{-S_{N}(V) + s_{\uN}\psi_{\uN}-\hat S_{N,K}[\psi](V,v,\Phi,w,\bar w,\eta,\bar\eta,\chi,\bar\chi)}\, .
\end{multline}
In the above formula, we have also included the standard gauge-fixing terms $s_{\uN}\psi_{\uN}$ for the factor $\uN$ of the gauge group $\uN\times\uK$ which is kept after the partial gauge-fixing. In particular, the partition function $Z_{N}$ of the $\uN$ gauge theory is given by
\be\label{ZZZZ} Z_{N} = \int\!\mathcal D V\mathcal D[\text{ghosts}]_{\uN}\, e^{-S_{N}(V)+s_{\uN}\psi_{\uN}}\, .\ee
Note that the D-brane probe action defined by \eqref{ANKdef1} does not depend on $\psi_{\uN}$ but \emph{does} depend on the equivariant gauge-fixing fermion $\psi$. 

Since, by construction, the integral over $w,\bar w,\eta,\bar\eta,\chi,\bar\chi$ is Gaussian, it can always be performed exactly and yields a functional superdeterminant
\be\label{pathI2} \Delta[\psi](V,v,\Phi) = \int\!\mathcal D w\mathcal D\bar w \mathcal D\eta\mathcal D\bar\eta\mathcal D\chi\mathcal D\bar\chi\,
e^{-\hat S_{N,K}[\psi](V,v,\Phi,w,\bar w,\eta,\bar\eta,\chi,\bar\chi)}\, .\ee
This determinant is automatically invariant under $\uN$, for any fixed values of $v$ and $\Phi$. Eq.\ \eqref{ANKdef1} is then equivalent to
\be\label{ANKdef2} \mathscr A_{N,K}[\psi](v,\Phi) = \Bigl(\frac{N}{K}+1\Bigr)S_{K}(v) -\ln\bigl\langle\Delta[\psi](V,v,\Phi)\bigr\rangle\, ,\ee
where the expectation value is taken in the ``background'' $\uN$ gauge theory. In the leading large $N$ approximation, \eqref{ANKdef2} simplifies to $\mathscr A_{N,K}[\psi]\simeq N A_{K}[\psi]$ and $\ln\langle\Delta[\psi]\rangle=\langle\ln\Delta[\psi]\rangle$, thus
\be\label{ANKdefN} A_{K}[\psi](v,\Phi) = \frac{1}{K}S_{K}(v) -\frac{1}{N}\bigl\langle\ln\Delta[\psi](V,v,\Phi)\bigr\rangle_{N=\infty}\, ,\ee
where only planar diagrams are taken into account to compute the expectation value $\langle\ln\Delta[\psi]\rangle_{N=\infty}$. 

Formulas for the gauge-fixing fermions $\psi$, the equivariant differential $\delta$, the gauge-fixing terms $\delta\psi$, etc., can be found in \cite{ferequiv}. These details are of course crucial for the study of specific examples, like in \cite{onemmbranes}, but they are not needed for our present general discussion and thus we shall refrain from copying them here. 

\subsection{Properties of the probe D-brane action}

The basic properties of the probe D-brane action listed in Section \ref{genDpSec} all follow straightforwardly from the above definition. In particular, the action is manifestly $\uK$ invariant and the form of the large $N$ expansion \eqref{AKNexp} follows directly from the discussion of Section \ref{diagsec}, which can be repeated without change, by simply including the ghosts into the list of vector-like variables.

The fundamental relations \eqref{fundrel} or \eqref{fundrel2} can also be derived along the lines of Section \ref{ZfromASec}, but now taking into account the effect of gauge-fixing. In particular, Eq.\ \eqref{ZNZNK} should be replaced by the more precise formula
\be\label{RelZZ}\boxed{\int\!\mathcal D v\mathcal D\Phi\mathcal D[\text{ghosts}]_{\uK}\, e^{-\mathscr A_{N,K}[\psi](v,\Phi) + s_{\uK}\psi_{\uK}}=\frac{Z_{N+K}}{Z_{N}}\,\cdotp}\ee
This can be derived from the crucial identity \eqref{SHdef}, by plugging the definition \eqref{pathI1} into the left-hand side of \eqref{RelZZ} and then using \eqref{stotfin}. Note that, even though the probe brane action itself does depend on the equivariant gauge-fixing fermion $\psi$ in a very non-trivial way, the path integral computed from it does not and yields the gauge-invariant ratio of partition functions, as expected. At large $N$, the path integral evaluates to the saddle point value $e^{-NA_{K}[\psi]^{*}}$ and, following the reasoning in Section \ref{ZfromASec}, we get \eqref{fundrel} and \eqref{fundrel2}: \emph{the planar free energy is equal to half the on-shell value of the D-brane probe effective action,} or, equivalently, using the leading term in the expansion \eqref{ZNexp},
\be\label{fundrel3} F^{(0)} = \frac{1}{2K}A_{K}[\psi]^{*}= \frac{1}{2}A[\psi]^{*}=\frac{1}{2K}A_{K}^{*}=\frac{1}{2}A^{*}\, .\ee
In particular, the leading large $N$ probe D-brane action $A_{K}$ does not depend on the equivariant gauge-fixing fermion $\psi$ when evaluated on-shell.

This fundamental property is ensured by the general formalism, but checking it explicitly is quite non-trivial and spectacular, even in the one matrix model. The explicit mechanism at work, ensuring the gauge-invariance, involves in a crucial way the quartic ghost terms, which induce a solution of the equations of motion for which ghost condensation occurs and the equivariant BRST symmetry is spontaneously broken \cite{onemmbranes}. This solution corresponds to the mininum of the D-brane action and the corresponding critical value of the action can be shown explicitly to be independent of the gauge fixing \cite{onemmbranes}. Ghost condensation, the associated spontaneous breaking of the equivariant BRST symmetry and its close relation with gauge invariance are expected to be quite generic phenomena and to also occur in models like the pure Yang-Mills theory.

\subsection{Remark on the bare bubble approximation}

The bare bubble approximation described in Section \ref{bareapsec} amounts to replacing the exact large $N$ formula \eqref{ANKdefN} by
\be\label{ANKbare} A_{K}^{\text{b.b.}}[\psi](v,\Phi) = \frac{1}{K}S_{K}(v) - \frac{1}{N}\ln\Delta[\psi](V_{\text{cl}},v,\Phi)\, ,\ee
where $V_{\text{cl}}$ is the classical value of the fields living on the background branes in the state under consideration. The discussion of Section \ref{bareapsec}, where the interest and meaning of this approximation was discussed at length, is unchanged by the addition of the gauge-fixing terms. However, the gauge-fixing procedure brings in a new important subtlety. 

Since the D-brane action computed from \eqref{ANKbare} is not exact, nothing ensures a priori that its on-shell value $(A_{K}^{\text{b.b.}}[\psi])^{*}$ be gauge-independent. This is to be contrasted with the situation in more familiar approximation schemes, like perturbation theory, which are expansions in terms of a small parameter. The gauge invariance of the exact answer then automatically ensures the gauge invariance order by order in the expansion. The bare bubble approximation is not an expansion in terms of a small parameter. This is actually its strength: it is non perturbative and does not rely in the existence of a small parameter. The price to pay is the gauge-dependence of the approximate result. Let us note that explicit calculations in the one matrix model, using a general $\xi$-gauge in which $\psi$ depends on an arbitrary parameter $\xi$, confirms that $(A_{K}^{\text{b.b.}}[\psi])^{*}$ does indeed depend on $\xi$ (whereas the on-shell value of the exact result does not). 

This may be seen as a flow, but we believe that it is not and should rather be viewed as a interesting and rich aspect of the approximation. It means that we really have a different bare bubble approximation for each gauge choice, and thus we may be able to adjust our gauge choice to improve the reliability of the approximation. Moreover, the ability of the approximation to capture the fundamental qualitative properties of the models, like for example the mass gap property or confinement, are not expected to depend sharply on the choice of gauge, at least within a reasonable class. For example, in the one matrix model, one finds that the bare bubble free energy, computed from \eqref{ANKbare} and \eqref{fundrel3}, does depend on the gauge-fixing parameter $\xi$. However, we get an excellent approximation for a wide range of values of $\xi$. Moreover, the most natural values, like $\xi=0$ (the Landau gauge), or another value of $\xi$ which is chosen such that the approximation and the exact result match at two loops, turn out to be extremely reliable.   

\subsection{Closed string gauge symmetries and equivariant gauge fixing}

From a standard string theory point of view, the non-abelian D-brane action can in principle be computed by evaluating open string disk diagrams attached to the D-branes with the insertion of closed string vertex operators describing the coupling to the non-trivial emergent holographic background. The result should be unambiguous, up to the action of closed string gauge symmetries. In the supergravity limit, this includes bulk diffeomorphisms, Neveu-Schwarz and Ramond-Ramond forms gauge symmetries. How these closed string gauge symmetries can be implemented quite generally in a non-abelian D-brane action was discussed at length in \cite{fer2}. On the other hand, from the point of view of our gauge-theoretic construction of the D-brane action, the only ambiguity comes from the choice of the equivariant gauge-fixing condition. It is thus natural to guess that the closed string gauge invariances should be related to the choice of an equivariant gauge-fixing fermion. Let us note that a philosophically similar remark was made in \cite{insightbrane} and that the seed of this idea could be traced back to \cite{Yone}. Further discussion of the physical consequences of the gauge-dependence of the brane action will be provided in the next Section.

\section{Discussion\label{dissec}}

In this Section, we wish to discuss some of the physical consequences of our construction of the D-brane probe action, especially in relation with the gauge-dependence of the framework. Our main point concerns the problem of bulk locality. We also draw some interesting analogies with the 't~Hooft Abelian projection ideas.

\subsection{On bulk locality}

\subsubsection*{Two problems which look superficially very different}

We would like to propose a relation between the following two superficially very different problems.

\smallskip

\noindent\emph{Problem 1: bulk locality} 

As already briefly reviewed in Section \ref{IntroSec}, this problem is notoriously difficult. Many deep questions in quantum gravity, including the physics of black holes, seem to be related to the understanding of bulk locality and its breakdown. The main difficulty stems from the fact that local observables are not gauge invariant. As a consequence, the notion of space-time locality cannot be a precise concept in quantum gravity. However, it is clearly crucial to understand how local observations can make sense at least approximately. An interesting point of view is based on the idea that locality can be recovered in relation with a reference background, but this reference background must itself fluctuate and back-react on the quantum gravitational physics. In the context of holography, it is not clear, to my knowledge, how one can ``decode'' the hologram from this point of view. In all cases, a simple physical picture allowing to understand how the familiar locality could emerge approximately in spite of the fact that it is not a gauge-invariant concept would be most welcome.

\smallskip

\noindent\emph{Problem 2: low energy physics in GUT models}

Consider a grand unified model with gauge group $G$ which is broken down to a ``standard model'' gauge group $H\subset G$ by the Brout-Englert-Higgs mechanism. Let $M$ be the grand unification mass scale and $m$ the typical scale associated with the standard model physics accessible to experiments. When $M\gg m$, we can clearly use an approximate description in which only the standard model is kept. This is a very familiar approximation. Yet it comes with a subtle facet which is usually overlooked or not discussed.

The basic justification of the approximation comes from the low energy effective action, or renormalization group, idea. Describing the physics by using the standard model degrees of freedom only, one makes an error which becomes arbitrarily small when $m/M\rightarrow 0$. The idea is completely general and applies to any quantum field theory with a separation of mass scales between degrees of freedom.

However, an additional subtlety shows up when the model is a gauge theory. When one works within the standard model, one uses observables that are invariant under the standard model gauge group $H$. But these are not gauge invariant under the gauge group $G$ of the fundamental theory and as such are not genuine observables! This implies that the standard model description is not simply about computing well-defined observables in a low energy approximation; it is also about using ``observables'' that are actually non-gauge invariant and thus only approximately physical, or ``fuzzy,'' in the sense that they are not uniquely determined in the fundamental theory. In particular, if the ratio $m/M$ is increased, it is not only the accuracy of the low energy description which will become questionable, but, more deeply, the physical meaning of the standard model observables will be lost. 

\medskip

There is a clear mathematical analogy between the two problems that we have just outlined. In both cases, the physics is expected to be naturally described, in some regime, by fuzzy, non gauge-invariant observables: the local observables in quantum gravity for the problem 1; the $H$-invariant but not $G$-invariant observables for the problem 2. In the case of the problem 1, the physics underlying the emergence of the approximate local description is unclear; in the case of the problem 2, the physical picture is much more familiar. We would like to propose that this relation between the two problems is actually more than a simple mathematical analogy but could be used as a basic physical picture for understanding bulk locality in general holographic contexts. 

\subsubsection*{The point of view of D-brane probes}

Let us see how the above idea incarnates from the point of view of D-brane probes. 

First, the gauge ambiguity in the definition of the probe action discussed in Section \ref{gfsec} is intimately related to the impossibility to build a gauge invariant local description of the emerging bulk holographic geometry. The bulk coordinates, for example the variable \eqref{phiYMdef} in the pure Yang-Mills theory or the $K\times K$ lower-right corners in the decomposition \eqref{funddec} of the elementary scalars in the $\nn=4$ theory, are not gauge invariant in the full, $\text{U}(N+K)$ invariant, microscopic description. This is perfectly consistent with our expectations, including the general remarks made in Section \ref{IntroSec}. It is impossible to define local quantities in the bulk, like the coordinates, in terms of gauge-invariant objects in the microscopic theory.  

Second, the bulk coordinates $\phi$, which are scalars from the point of view of the probe brane worldvolume, enter in the microscopic description in a rather specific way. They are always associated with ``Higgs-like'' terms giving masses to the off-diagonal components $w$, $\bar w$ which include the ``W-like'' bosons, see e.g.\ Eq.\ \eqref{tildeSmYM}. We are using quotation marks because, as we have already emphasized before, the motion of the D-brane probes in the holographic space transverse to the background branes is not associated to a usual Higgs mechanism in general. However, the way the coordinates enter and the fact that a non-zero $\phi$ corresponds to giving a mass to the off-diagonal components of the gauge field imply that there is always at least a formal analogy, whose physical meaning will be further clarified in the next subsection when we discuss the Abelian projection. 

In particular, at large $\phi$, which corresponds to the UV or the near-boundary region of the holographic bulk, the off-diagonal fields will be very heavy. In this region, we thus expect the gauge dependence of the construction to be essentially irrelevant and bulk locality to be meaningful, for the very same reasons as the standard model description of the low energy physics is accurate when $M\gg m$. On the other hand, deep in the bulk, which is the IR region, the notion of locality will become fuzzy, imprecise, the associated local quantities becoming strongly gauge-dependent, in the same way as the observables in the standard model would be fuzzy and imprecise without a sharp separation of scales between $M$ and $m$.

\subsubsection*{Gauge fuzziness, locality, space-time uncertainty, complementarity}

The fuzziness of space we have just discussed is not the same kind of fuzziness which is usually associated with quantum fluctuations of space-time and quantum gravity effects at the Planck length. In our context, these  are related to the $1/N$ corrections, which play no r\^ole in the present discussion. We are dealing here with a new sort of fuzziness, and it is likely that it could be used to reinterpret, or shed some new light on, many phenomena in gravity. Working out its full consequences is of course far beyond our present goal. Let us however discuss the simplest case: a D3-brane probing $\text{AdS}_{5}\times\text{S}^{5}$ \cite{Douglas0}.

As already reviewed in Section \ref{spacedimsec}, the transverse dimensions are then the six scalars $\phi_{m}$ in the set of variables $v$ appearing in the decomposition \eqref{funddec} of the $\nn=4$ super Yang-Mills elementary fields. Because of supersymmetry, the brane feels no potential; even on-shell, it can thus sit anywhere in the six transverse dimensions. This ambiguity is associated with a choice of a particular point on the branch of the moduli space of the $\nn=4$ theory corresponding to a Higgsing of $\text{U}(N+1)$ down to $\uN\times\text{U}(1)$ (or, more precisely, of $\text{SU}(N+1)$ down to $(\uN\times\text{U}(1))/\text{U}(1)$ since it is well-known that the overall $\text{U}(1)$ is decoupled in the AdS description). Because of conformal invariance, the location of the brane is not physical. It simply sets a ``GUT'' mass scale $M\sim \sqrt{\phi_{m}\phi_{m}}\sim r/\ls^{2}$, where $r$ is the usual radial coordinate in AdS.

What does our above discussion relating the ``fuzziness'' of space-time to the gauge-dependence tells us in this context? Let $E$ by the typical energy of an excitation that appears on the probe brane. The location of the brane should be well-defined as long as $E$ is much smaller than the ``GUT'' scale $r/\ls^{2}$, but should become fuzzy if $E\sim r/\ls^{2}$. This is exactly what is found in explicit calculations \cite{Douglas1}! Moreover, combining this result with the quantum mechanical uncertainty relation $\Delta t\Delta E\gtrsim 1$, we make the link with Yoneya space-time uncertainty relation $\Delta t\Delta r\gtrsim \ls^{2}$ \cite{spacetimeun}.

It would be very interesting to try to extend this kind of reasoning in the presence of a horizon in the bulk geometry. Could it be that two different gauge choices yield two completely different-looking, albeit strictly equivalent, description of the physics near the horizon? This statement looks like a mathematically precise notion of ``complementarity'' for black holes \cite{complementarity}. We might even want to speculate further. We have seemingly contradictory descriptions of black hole physics: one, given by an observer at infinity, in which unitarity is made extremely hard to understand by the Hawking evaporation process; and another, given by an observer crossing the horizon. It is then tempting to ask the following question. Could it be that these two drastically different physical picture are actually related by a transformation akin to the transformation between a renormalizable gauge in gauge theory, in which the UV behavior is very simple but unitarity deeply hidden by the propagation of ghosts, and a non-renormalizable but unitary gauge, with no propagating ghosts, for which unitarity is manifest? Clearly these are only hypothesis at the present stage, which might be suggested by our approach, but which require much more study before they can be established.

\subsection{On the Abelian projection scenario\label{APSec}}

Let us now point out some very interesting similarities between our D-brane probe approach and the 't~Hooft Abelian projection, which originated in \cite{tHooftAP} and has been much studied since then, in particular in numerical lattice calculations; see e.g.\ \cite{APreview} for a review.

The Abelian projection is a reformulation of the pure Yang-Mills theory, or of QCD, as an Abelian gauge theory, in which, hopefully, the low energy properties are easier to understand than in the original, non-abelian formulation. It is equivalent to a partial fixing of the gauge group from $\uN$ down to $\text{U}(1)^{N}$ (or from $\text{SU}(N)$ down to $\text{U}(1)^{N-1}$; the global $\text{U}(1)$ in $\uN$ decouples). 't~Hooft noticed that the gauge-fixing conditions breaking $\uN$ down to $\text{U}(1)^{N}$ can have monopole-like singularities along worldlines in spacetime, corresponding to the enhancement of the $\text{U}(1)^{N}$ gauge symmetry to a non-abelian group. This suggests that a correct treatment of the Abelian projected model should include these monopoles explicitly. The main interest of this description is that all the ingredients for the dual superconductor picture of confinement \cite{dualsupra} are explicitly present. Confinement of electric charges would automatically follow from the condensation of the monopoles, if it occurs.

In spite of the very nice physical picture it offers, there are several well-known difficulties associated with the Abelian projection scenario. The Abelian reformulation with monopoles is hard to derive rigorously from first principles and is a priori strongly quantum mechanical. Using this formulation to get an analytic understanding of the low energy physics, for example to prove monopole condensation, doesn't seem very promising. An important ingredient allowing to simplify the picture is the idea of Abelian dominance \cite{ADominance}, which states that the non-abelian degrees of freedom become massive and can thus, in some approximation, be discarded at low energy. Unfortunately, proving Abelian dominance seems, at first sight, as difficult as proving the mass gap property in Yang-Mills. For all these reasons, most of the work done in the Abelian projection context has focused on numerical studies on the lattice (see, however, \cite{kond} and references therein). Another seeming drawback of the approach, which is often emphasized, is the strong gauge dependence of the whole construction, including of the fundamental concept of monopole.

From the point of view of the present paper, it is clear that the Abelian projection is equivalent to a picture where a stack of $N$ branes classically sitting on top of each other is treated as a collection of $N$ individual branes, with gauge group $\text{U}(1)^{N}$. In particular, the correct partial gauge fixing method that one must use is the procedure reviewed in Section \ref{equivsec}. This brane interpretation of the Abelian projection allows to make an analogy with our own construction. Of course, the D-brane probe story has some major differences: the partial fixing $\text{U}(N+K)\rightarrow\uN\times\uK\rightarrow\uK$ is not quite the Abelian projection $\uN\rightarrow\text{U}(1)^{N}$; more importantly, unlike for the Abelian projection, the large $N$ limit is very natural in our framework. However, the analogy allows to infer some interesting and potentially fruitful and enlightening relations between ideas, concepts and methods developed quite independently and without obvious relations to each other.

For example, the gauge dependence of the Abelian projection constructions is now seen as being strictly analogous to the gauge dependence of the notion of a local holographic bulk spacetime. From this perspective, the gauge dependence is no longer a flaw, but rather a necessary property of any consistent holographic framework useful to deal with the strong coupling properties of a gauge theory. Of course, the on-shell physics is gauge independent, but the simplest and most natural picture of the physics is obtained in a gauge-dependent formulation. 

Another drawback of the usual Abelian projection idea is also solved in our point of view: the probe brane effective action is automatically classical at large $N$ and thus we no longer have to worry about possible strong quantum effects. This is a property of the large $N$ limit for a probe and is thus not seen in the usual $\text{U}(1)^{N}$ framework. 

The idea of Abelian dominance also has a very simple holographic interpretation:  the emergent dimension(s) play(s) the r\^ole of the scalar field(s) allowing the ``Higgsing'' giving mass to the non-abelian, or off-diagonal, degrees of freedom. Actually, the gluon and ghost condensates which are commonly used to study Abelian dominance (see e.g.\ \cite{Schaden,schaKAD} and references therein) are precise analogues of the bilinears \eqref{phigenform} associated with the holographic space! In the holographic description, the D-brane probes will feel a potential in the bulk space and the non-zero values of these condensates correspond to the equilibrium position $\phi\sim\La^{2}$ of the D-brane probes sitting at the minimum of the potential.

In conclusion, our D-brane picture sheds an interesting light on some aspects of the Abelian projection philosophy. Conversely, we expect that ideas developed by studying the Abelian projection scenario will be useful to work out the holographic physics of D-brane probes in general contexts, like the pure Yang-Mills theory. In particular, the numerical lattice techniques may be of great help.

\section{Conclusions and outlook}

Let us summarize our main results:

(i) We have proposed to extend the usual string-theoretic notion of D-branes to any gauge theory, like the pure Yang-Mills theory. This allows to define from first principles the notion of probe D-branes in the presence of background D-branes on which the gauge theory lives in contexts much wider that those that are traditionally considered in the literature, while keeping some crucial intuitions from D-brane physics in string theory, in particular their use to understand holography. Moreover, the probe D-brane action captures the full large $N$ physics of the original theory and thus provides a new formulation of the large $N$ limit of gauge theories.

(ii) In the limit where the number of background branes is much larger than the number of probe branes, we have shown how the mixed vector/matrix model structure of the Feynman diagrams computing the D-brane probe action naturally lead to an effective action describing the motion of the probes in an emerging holographic space-time sourced by the background branes. More generally, we have shown how to define the open string field theory, containing an infinite number of excited open string states, living on the D-branes probing the holographic geometry.

(iii) We have described a non-perturbative approximation scheme, the ``bare bubble approximation,'' which can be used in all models, even when there is no supersymmetry. The leading order of this approximation already resums an infinite set of loop diagrams, corresponding to worldsheets of all topologies. One can thus obtain analytical non-perturbative information about the theories. We have explained how this approximation can be improved, in particular using numerical lattice calculations.

(iv) We have explained in details that the D-brane probe dynamics and in particular the local bulk physics seen by the probes depends on a choice of partial gauge-fixing, which is a crucial piece in the definition of the D-brane probe action. This gauge-fixing is governed by an equivariant version of the usual BRST cohomology and parameterized by an equivariant gauge-fixing fermion $\psi$. The gauge-fixed Lagrangian must contain quartic ghost terms at tree level. These terms can induce ghost condensation and the spontaneous breaking of the equivariant BRST symmetry. The resulting on-shell physics, obtained by solving the classical equations of motion for the D-brane probes, is always gauge-invariant.

(v) The notion of bulk locality becomes fuzzy deep in the bulk, in the same way as the low energy description of the physics by the standard model would become fuzzy if the grand unification scale were lowered to the typical standard model energies; local coordinates in the bulk are not physical because they are not diffeomorphism invariant, whereas standard model observables are not physical because they are not invariant under the grand unified gauge group. The D-brane actions, although they may provide very different bulk spacetime pictures, are ensured to be strictly equivalent on-shell. This mechanism might yield a precise mathematical implementation of the idea of black hole complementarity. 

(vi) We have emphasized the close relation between the D-brane probe picture and some aspects of the Abelian projection scenario, yielding new points of view on the problem of the gauge-dependence of this approach as well as on the notion of Abelian dominance. In particular, analogues of the gluons and ghosts condensates that have been heavily studied in this context correspond to the emerging holographic dimensions in the D-brane probe picture. This opens the way to transpose the numerous techniques, both analytic and numerical, that have been developed in the study of the Abelian projection, to the context of holography in real-world theories like QCD.

Clearly, much remains to be done to fully comprehend the consequences of the above constructions and claims. Let us briefly mention a few natural avenues of future research.

First, it seems natural to study some very simple exactly solvable set-ups, 
on which several aspects of our formalism, especially the validity of the bare bubble approximation, can be explored and tested. The simplest non-trivial example is the one matrix model \cite{onemmbranes} and there are several other cases that would be worth studying, from the matrix quantum mechanics, which is related to two-dimensional string theory, to the quantum mechanics of D0-branes and the associated black hole at finite temperature.

On the front of exact analytic results in supersymmetric contexts, our microscopic construction of D-brane probe actions provides in principle an  arena where the many powerful techniques developed over the last two decades (holomorphy, localization, matrix model, etc., see e.g.\ \cite{exacttech} and references therein) could be applied. An important first step in this direction is to develop a supersymmetric version of the equivariant gauge-fixing procedure explained in \cite{ferequiv}. 

Of course, the most exciting applications concern models for which a holographic description has not already been found and studied using other methods. There are interesting examples in the supersymmetric realm, as the famous $\nn=2$, $N_{\text{f}}=2N$ superconformal Yang-Mills theory \cite{rastelli}, but of course the most salient cases are non-supersymmetric models like the pure Yang-Mills theory. On the analytical side, it is urgent to investigate the physics captured by the bare bubble approximation for such models, whereas on the numerical side, lattice techniques seem promising, as we have explained.

Finally, let us note that the framework is not limited to ordinary gauge field theories living on the branes. One may start with an open string field theory, for example the open string field theory living on D3 branes in type IIB string theory, and study the resulting probe brane action exactly as we have done before. In principle, this can yield holographic spacetimes which can be asymptotically flat, since the near-horizon or low energy limit is never used.

\subsection*{Acknowledgments}

Important pieces of this work have been presented in seminars given at the international conference ``Quantum Gravity in Paris'' (March 2013), at CERN, the University of Pisa, the University of Surrey, Imperial College (May 2013) and the XLIIIth Summer Institute of the Laboratoire de Physique Th\'eorique de l'\'Ecole Normale Sup\'erieure (August 2013). I have greatly benefited from the many questions and the long and enthralling discussions that most often followed these talks. I also apologize to all the people attending the talks for always promising that the paper would appear ``next week.'' I also wish to thank particularly M.~Mari{\~n}o, C.~N\'u{\~n}ez, M.~Petrini, B.~Pioline, J.~Troost and my students M.~Moskovic and A.~Rovai for very useful exchanges.

This work is supported in part by the belgian FRFC (grant 2.4655.07) and IISN (grant 4.4511.06 and 4.4514.08).

%
%
%

%

%

%
\end{document}